\documentclass[12pt,fleqn]{article}%
\usepackage{eurosym}
\usepackage{amsfonts,amsmath,amssymb,graphicx,setspace}
\usepackage[usenames, dvipsnames]{xcolor}
\usepackage[colorlinks=true,linkcolor=blue,citecolor=Mahogany,urlcolor=blue]%
{hyperref}
\usepackage[top=1in, bottom=1in, left=1in, right=1in]{geometry}
\usepackage[normalem]{ulem}
\usepackage{subfig}
\usepackage{amsmath}
\usepackage{amsfonts}
\usepackage{amssymb}
\usepackage{amsthm}
\usepackage{mathtools}
\usepackage{graphicx}
\usepackage[longnamesfirst]{natbib}
\usepackage[normalem]{ulem}
\definecolor{accentcolor}{RGB}{26, 148, 49}
\usepackage{hyperref}
\hypersetup{pdfborderstyle={},allcolors=accentcolor,colorlinks}

\usepackage{titling}
\usepackage{bbm}
\usepackage{apptools}

\usepackage{etoolbox}
\newcommand{\zerodisplayskips}{%
  \setlength{\abovedisplayskip}{8pt}%
  \setlength{\belowdisplayskip}{7pt}%
  \setlength{\abovedisplayshortskip}{8pt}%
  \setlength{\belowdisplayshortskip}{7pt}}
\appto{\normalsize}{\zerodisplayskips}
\appto{\small}{\zerodisplayskips}
\appto{\footnotesize}{\zerodisplayskips}

\AtAppendix{\counterwithin{lemma}{section}}
\AtAppendix{\counterwithin{claim}{section}}
\usepackage[
]{ifdraft}

\usepackage[font=normalsize,justification=centering]{caption}
\DeclareCaptionLabelFormat{captionf}{\textsc{#1~#2}}
\usepackage{lscape}
\usepackage{placeins}
\providecommand{\U}[1]{\protect\rule{.1in}{.1in}}
\usepackage[noamsmath]{kpfonts}
\usepackage{inconsolata}

\usepackage[T1]{fontenc}
\providecommand{\U}[1]{\protect\rule{.1in}{.1in}}

\newtheorem{proposition}{Proposition}
\newtheorem{corollary}{Corollary}
\newtheorem{lemma}{Lemma}

\newtheorem{claim}{Claim}

\begin{document}

\title{Collective Progress: Dynamics of Exit Waves\thanks{We thank
Arjada Bardhi, Alessandro Bonatti, Francesc Dilme, Hari Govindan, Faruk Gul, Toomas Hinnosaar, Alessandro Lizzeri, Chiara Margaria, Pietro Ortoleva, Wolfgang Pesendorfer, and Heikki Rantakari for helpful comments and suggestions. We also thank seminar
audiences at Central European University, Collegio Carlo Alberto, Princeton, Seoul National University, Tel-Aviv University, UCLA, University of Rochester, and VSET. Yariv
gratefully acknowledges financial support from the National Science Foundation
through grants SES-1629613 and SES-1949381.}}
\author{Doruk Cetemen\thanks{Collegio Carlo Alberto;
\href{mailto:doruk.cetemen@carloalberto.org}{\texttt{doruk.cetemen@carloalberto.org}%
}} \hspace{3mm} \hspace{2mm} Can Urgun\thanks{Princeton University;
\href{mailto:curgun@princeton.edu}{\texttt{curgun@princeton.edu}}}
\hspace{3mm}\hspace{2mm} Leeat Yariv\thanks{Princeton University, CEPR, and
NBER; \href{mailto:lyariv@princeton.edu}{\texttt{lyariv@princeton.edu}}}}
\date{\today }

\pretitle{\begin{flushleft}\LARGE} 
\posttitle{\end{flushleft}}
\preauthor{\begin{flushleft}\large} 
\postauthor{\end{flushleft}}
\predate{\begin{flushleft}} 
\postdate{\end{flushleft}}
\settowidth{\thanksmarkwidth}{*}
\setlength{\thanksmargin}{-\thanksmarkwidth}

\maketitle
\thispagestyle{empty}

\renewenvironment{abstract}
 {\par\noindent\textbf{\abstractname.}\ \ignorespaces}
 {\par\medskip}
\begin{abstract}
We study a model of collective search by teams. Discoveries beget discoveries and correlated search results are governed by a Brownian path. Search results' variation at any point---the search scope---is jointly controlled. Agents individually choose when to cease search and implement their best discovery. We characterize equilibrium and optimal policies. Search scope is constant and independent of search outcomes as long as no member leaves. It declines after departures. A simple drawdown stopping boundary governs each agent’s search termination. We show the emergence of endogenous exit waves, whereby possibly heterogeneous agents cease search all at once.

\vspace{2mm}

 \vspace{2mm}


\end{abstract}

\newpage
\pagenumbering{arabic}

\section{Introduction}

Discoveries are often made by teams. Advances in motor vehicles, communication devices, and pharmaceuticals frequently take place as joint ventures. Understanding collective progress is therefore vital for the analysis of innovation. Much of the literature on teamwork has focused on experimentation models, starting from the canonical work of \cite{bolton1999strategic} and \cite{keller2005strategic}. Those models center on teams' efforts to ascertain whether one direction or project is superior to another. Nonetheless, many discovery processes follow a path of search. Building on past discoveries, teams come up with new ones. Furthermore, there is a richness of dynamics in collective efforts not captured in prior models---alliances tend to dissolve over time, with exiting members exploiting knowledge accrued during their collaborations.


This paper offers a new framework for studying collective progress based on a process of search. We identify how the breadth of search and decisions to terminate search vary with members' characteristics, the synergies in place. We also show that exit waves, where multiple members halt search simultaneously, are an inherent feature of such processes.

Technological developments rarely occur in a vacuum and discoveries build on
one another. We therefore consider environments in which search results are
correlated over time and follow a Brownian path, as first modeled by
\cite{callander2011searching}. The scope of search, captured by the Brownian
path's instantaneous variance, is chosen at each moment by the searching alliance.
Specifically, each member of a searching alliance incurs a strictly
positive cost that depends on that member's own search scope. While jointly searching,
the Brownian path's instantaneous variance corresponds to the
sum of members' search scopes. Any member can terminate her search at any
point. A member ceasing her search receives a lump sum payoff corresponding to
the maximal value the search has produced till her departure. 
Certainly, some alliance members may choose to continue their search even after other
members have exited. These remaining members experience prolonged search costs, but
benefit from any further breakthroughs, as reflected by search results that
exceed the previously-observed maxima. As search progresses, members gradually
terminate their search until it halts altogether.\footnote{Most of our qualitative results carry over when introducing penalties for later exits, though naturally such penalties alter exit patterns. In particular, penalties for later exits can introduce exit waves mechanically---once one agent departs, others may follow suit to avoid penalties.}

We characterize equilibrium search in Markov strategies, where
state variables correspond to the current search results, the attained
maximum, and the active alliance. We show that, in any active alliance, search scope is constant and
independent of search results as long as no member leaves. Individual search scopes increase when members depart, reflecting the more limited
free-riding opportunities present. The optimal time at which members depart
and alliances shrink is governed by a simple stopping boundary, often referred
to as a drawdown stopping boundary. Such boundaries are defined by
one number, the drawdown size. Whenever search results fall by more
than the drawdown size relative to the maximal
observation achieved, a subset of members ceases search.

The ratio of marginal to fixed costs governs both equilibrium search scopes and drawdown sizes and serves as a proxy for the synergies present in an alliance. In particular, agents may prefer to team up with others exhibiting both higher marginal and fixed costs, provided the ratio guarantees they are more willing to contribute to the collective search.

Relative to an individual searching on her own, standard free-riding motives
drive search scopes down in an alliance. This is a form of a discouragement
effect, whereby members do not search as intensely when they expect others to
bear some of the search costs. Nonetheless, externalities make
search more valuable in a team: a member can reap
the benefits of her peers' efforts. There is therefore also an
encouragement effect, reminiscent of that present in experimentation settings, that leads team members to search for longer than they
would have on their own.

Our equilibrium characterization allows us to identify members' patterns of exits. In general, those
exhibiting high ratios of marginal to fixed costs leave earlier
than those exhibiting low such ratios. We show that, even when individual
costs are fully heterogenous, clustered exits, or exit waves, may occur in
equilibrium. Importantly, while the precise timing of exit waves may depend on
the realized path of discoveries, their sequencing---who leaves first, second,
etc., and with whom---does not.

Beyond its substantive implications, our equilibrium characterization offers a technical contribution. As we detail in our literature review below, extant analyses of single-agent search processes often resort to modeling short-lived agents, absent any controls. In contrast, we analyze the evolution of collective search by forward-looking and sophisticated agents who can utilize a costly control.

In the last part of the paper, we
characterize the socially optimal search scope and stopping policies. The
socially optimal search scope is also constant and independent of search
results within any active alliance. Naturally, the positive
externalities induced by each member's investment in search scope imply that
the socially optimal level is higher than that chosen in equilibrium.
Furthermore, in contrast to equilibrium search scopes, as alliance members
terminate their search, the optimal scope of those remaining declines. Optimal exits are
governed by drawdown stopping boundaries, although the drawdown sizes
corresponding to each active alliance differ from those determined in
equilibrium---optimal drawdown sizes are larger, corresponding to longer
search durations.\footnote{As we show, allowing for non-Markovian equilibria does not eliminate some of the inefficiencies we highlight.} In terms of exit waves, clustered exits may be optimal even when individuals incur fully
heterogeneous costs. As in equilibrium, the sequence of optimal exit waves is deterministic and
independent of the realized search path. However, optimal exit waves may differ substantially from those induced in equilibrium.

Finding the optimal sequence of exit waves is a challenging combinatorial
problem. A social planner needs to consider all possible ordered partitions of the
original searching team and assess search outcomes from the corresponding exit
wave sequences. We show a simple method for identifying the
optimal sequencing for one class of settings, when individual search costs
are proportional to one another. Similar to equilibrium, the social planner
terminates the search of those with the highest search costs first. This
limits the exit wave sequences to consider. We illustrate a
simple procedure, akin to a greedy algorithm (see, e.g.,
\citealp{papadimitriou1998combinatorial}) that yields the optimal exit wave
sequence. In rough terms, the social planner can use a recursive procedure,
first identifying the optimal last alliance to search---the alliance
that would generate the highest welfare when all members are constrained to
stop jointly. Once that alliance is identified, the social planner can find
the optimal penultimate alliance. And so on. The procedure allows us to
highlight settings in which equilibrium exit waves differ substantially from
those set optimally.

\section{Literature Review}
Since \cite{weitzman1979optimal}, much of the search with recall literature
has focused on individual agents' discovery process, where the set of options
is independent of one another. Our consideration of a Brownian path of
discoveries, capturing intertemporal correlations, is inspired by the setting
studied by \cite{callander2011searching}. He studies short-lived agents who
decide whether to choose an optimal, previously explored, result or experiment
on their own. Most of the work that ensued considers behavior of short-lived agents as well. \cite{urgunyariv2020} analyze an individual-search setting similar
to the one analyzed here. 

In recent years, substantial attention has been dedicated to the study of
collective experimentation. Much of this literature focuses on
learning spillovers between team members. For instance, the classic papers of
\cite{bolton1999strategic}, \cite{keller2005strategic} extend the
two-armed bandit problem to a team setting, where agents learn from
others. Information is a public good. Thus, there is a free-rider problem that
discourages experimentation. Nonetheless, there may also be an encouragement
effect through the prospect of others' future experimentation. See
\cite{horner2016learning} for a survey.

Another strand of literature inspects settings in which stopping is determined collectively.  \cite{albrecht2010search} and \cite{strulovici2010learning} consider sequential search and experimentation,
respectively, where a committee votes on when to stop. They illustrate when
collective dynamics may impede search or experimentation.
\cite{bonatti2016politics} offer a model in which agents exert effort on
different projects but stop experimentation jointly. Optimally, one agent advances her preferred project quickly. Her opponent
agrees to early advanced projects in order to limit effort. 
\cite{deb2020fostering} take a design
perspective---for a given deadline at which a project has to be chosen, the
principal commits to a selection rule. \cite{titova2019collaborative} studies
a public-good setting in which a team decides whether to implement a
public good. Payoffs are revealed through a Pandora's box problem \`{a} la
\cite{weitzman1979optimal}. Optimal
information and projects are selected, but free-riding may generate
inefficient delays.\footnote{Dynamic contribution games without
experimentation or uncertainty have also been heavily studied, see for
instance \cite{admati1991joint}, \cite{marx2000dynamic},
\cite{yildirim2006getting}, and  \cite{cetemen2019uncertainty}.}


There are also several papers illustrating patterns reminiscent of the
clustered exits we characterize, mostly in settings in which agents have
private information. \cite{bulow1994rational} consider a seller who dynamically reduces the price of identical goods until demand meets supply. Agents have independent valuations and decide if and when to buy. In equilibrium, frenzies, where multiple agents buy at the same price, may occur. \cite{caplin1994business} study a three-period
irreversible-investment game in which each firm receives private information
on the aggregate state of the economy as well as observes others' prior
decisions. Firms' actions reveal information and can generate a wave.
\cite{gul1995endogenous} analyze a two-agent model in which both try to
predict the value of a project using their private information. Each decides when to issue a prediction, where delay entails a
flow cost. The timing of decisions is then informative and clustered
predictions occur in equilibrium. \cite{rosenberg2007social} study a multi-agent
version of the standard real-options problem (see \citealp{dixit1994investment}).
Agents observe private signals about common returns to a risky project, as
well as the actions of others. If one agent switches to a safe
project---namely, exercises an option---this can lead the other agent to
immediately switch to the safe project as well. See also
\cite{murto2011learning} and \cite{anderson2017rushes}. In a static information-collection setting, \cite{bardhi2021local} characterize optimal subsets, or mini-publics, to be activated.\footnote{There is also
a literature that tries to explain industry \textquotedblleft
shakeouts,\textquotedblright\ corresponding to times at which firm numbers
plummet, absent a decline in output. For example, \cite{jovanovic1994life}
suggest shakeouts result from exogenous technological shocks. Initially,
firms enter new profitable markets. Profits decrease
as more firms enter. When there is a technological shock, some firms become
more productive than others, potentially leading to clustered exits.}

The techniques we develop relate to the applied mathematics
literature on optimal stopping, see \cite{peskir2006optimal} and \cite{azema}
for particularly relevant sources.

\section{A Model of Collective Search}

Consider a team of $N$ agents---product developers, academic researchers,
etc.---searching through a terrain of ideas in continuous time. Time is
indexed by $t$ and runs through $[0,\infty)$. Each seeks good outcomes and
ultimately benefits from the maximal value they have found when they stop
their search. Formally, we assume all agents are risk neutral. At each time
$t$, agent $i=1,2,...,N$ decides on the scope of her search $\sigma_{i,t}
^{A}\in\lbrack\underline{\sigma},\overline{\sigma}]$, where $\overline{\sigma}\geq\underline{\sigma}>0$ and $A\subseteq
\{1,...,N\}$ is the alliance of agents still searching at time $t$. Agents' scope choices are observed within their alliance.
As we soon describe, the search
scope naturally feeds into the breadth of search conducted by the alliance of
active agents. For $i=1,2,\ldots N$, any search scope $\sigma$ comes at a cost of $c_{i}(\sigma)$,
where $c_{i}$ is twice continuously differentiable, increasing, and convex, with a second derivative bounded above zero over $[\underline{\sigma},\overline{\sigma}]$. The special
case of $\underline{\sigma}=\overline{\sigma}$ corresponds to settings in which
search scope is not controlled and agents' only choose when to stop search.

We model the progress of discoveries using a Weiner process, which allows us
to capture the correlation of new developments over time, and the impact of
search scope of those who engage in search.\footnote{We view correlation as an
important feature of discovery processes. Nonetheless, from a purely
theoretical perspective, one could analyze an analogous model with independent
samples. As it turns out, such a model is far less tractable. Details appear in the Online Appendix.} Formally, for any time $t$, denote by
$B_{t}$ the standard Brownian motion with $B_{0}=0$, and let $\sigma_{t}^{A}$
denote the controlled breadth of search, which will depend on the search scopes
of all members of the active alliance $A$ as we soon describe. The observed
value at $t$---which can be thought of as the expected value of the discovery---is denoted by $X_{t}$, where $X_{0}=0$ and the law
of motion is given by:
\[
dX_{t}=\sigma_{t}^{A}dB_{t}\text{.}%
\]
Whenever the alliance $A$ of agents is searching, we assume $\sigma_{t}%
^{A}=\sum_{i\in A}\sigma_{i,t}^{A}$.\footnote{In the Online Appendix, we show
that our analysis can be directly extended to the case in which, for any alliance $A$,
we have $\sigma_{t}^{A}=f^{A}(\{\sigma_{i,t}^{A}\}_{i\in A})$, with $f^{A}$ a
differentiable function. Comparative statics would naturally depend on
alliances' technologies captured by $\{f^{A}\}_{A}$.}

The search scope can be interpreted in two ways. First, it can capture search
breadth. Investment in development, through acquisition of instruments or
expert time, often entails an increase in risk: it either leads to substantial
leaps, or to more pronounced losses. Second, given our modeling of search
values, the search scope can also be thought of as capturing search
\textit{speed}. Changing the search scope from $1$ to $\sigma$ at any small
interval of time is tantamount to \textquotedblleft speeding
up\textquotedblright\ the process by a factor of $\sigma^{2}$. As we soon
show, search \textit{returns} depend linearly on search scope.

We assume the discovery process exhibits no drift: in applications, the mere
passage of time rarely improves or worsens search outcomes over standard
horizons of research and development. Naturally, one could consider a team
that \textit{controls} drift rather than search scopes, which would also
translate to the returns of search with recall. The analysis would follow
similar lines to those we describe, although with an important loss in
tractability.\footnote{\cite{taylor1975stopped} characterize the maximal
value of search with constant drift. The resulting value is far less amenable
to further analysis than ours.} We view endogenous search scopes as natural for most
applications, where investments in innovation either affect the speed
at which progress is made, or entail non-trivial risks.

\subsection{Payoffs}

Each agent is rewarded according to the maximal project value
observed up to her stopping time. Let $M_{t}$ denote the maximum value
observed by time $t:$
\[
M_{t}=\left(  \max_{0\leq r\leq t}X_{r}\vee M_{0}\right)  ,
\]
where we assume that $M_{0}=0$.

For any aggregate fixed search scope $\sigma$, at time $t$, $\mathbb{E}%
(M_{t})=\sigma\sqrt{2t/\pi}$. Thus, the choice of search scope translates
directly to the expected returns from search.

When any agent $i$ stops at time $\tau$, her resulting payoff is given by
\[
M_{\tau}-\int_{0}^{\tau}c_{i}(\sigma_{i,t})dt\text{,}%
\]
where $\sigma_{i,t}$ is the timed search scope of individual $i$, which may depend on the alliances she is active in.\footnote{In Section \ref{LatePenalties}, we discuss an extension in which agents who stop later are penalized.} 
Any progress made after an agent stops searching does not
impact her payoffs.

Agents observe one another's search. In particular, whenever agents
stop searching, other agents realize their search will continue within a
smaller alliance. 

\subsection{Strategies and Equilibrium}

At any time $t$, the state of the environment is summarized by $X_{t},M_{t}$,
and $A_{t}$, where $A_{t}$ is the active alliance of agents still searching.

A strategy for agent $i$ dictates her chosen search scope over time and her
stopping policy. Formally, it is a pair of functions $(\sigma_{i}^{A},\tau
_{i}^{A})$, where $A\subseteq\{1,...,N\}$ and $i\in A$. In principle,
$(\sigma_{i}^{A},\tau_{i}^{A})$ may depend on time, as well as the entire path
of observed search values and corresponding maxima. Let $\{\mathcal{F}_{t}\}$ denote the natural filtration induced by the governing Brownian motion. Agents' strategies are adapted to this filtration.

We restrict attention to Markov strategies. That is, we assume agents use
strategies of the form $(\sigma_{i}^{A},\tau_{i}^{A})$ that depend only on the
state variables $X_{t}$, $M_{t}$, and $A_{t}$. Formally, $\sigma_{i}%
^{A}:\mathbb{R}^{2}\rightarrow\lbrack\underline{\sigma},\bar{\sigma}]$, and
$\tau_{i}^{A}$ is a random variable over $\mathbb{R}_{+}$ such that $\Pr
(\tau_{i}^{A}=t|\mathcal{F}_{t})=\Pr(\tau_{i}^{A}=t|X_{t},M_{t})$ for all $i$.\footnote{The inefficiencies we highlight do not vanish when considering equilibria in non-Markovian strategies, see our discussion in Section \ref{NonMarkovian}.}

We further assume that a continuous stopping boundary determines when each
agent halts her search. Formally, for all $i$ and all alliances $A$ such that
$i\in A$, the stopping policy takes the following form:
\[
\tau_{i}^{A}=\inf\{t\geq0:X_{t}=g_{i}^{A}(M_{t})\},
\]
where $g_{i}^{A}(\cdot)$ is a continuous function. This formulation implicitly
implies that, upon indifference, agents exit the search. Our assumption that stopping boundaries are continuous is without loss of generality as long as any agent is willing to search on her own, which we show in the Online Appendix. As we soon show, in our setting, departing agents would never benefit from continuing the search in a smaller alliance: the externalities offered by a larger alliance are always beneficial. 

Given $\left\{  (\sigma_{j}^{A},\tau_{j}^{A})\right\}  _{j\neq i}$, agent
$i$'s best-response strategy simply maximizes her expected payoff given this
profile. Formally, it is determined by solving the following problem for each alliance
$A$ such that $i\in A$:
\[
\sup_{\tau^{A},\left\{  \sigma_{i,t}^{A}\right\}  _{t=0}^{\tau_{i}}}%
\mathbb{E}_{\left\{  (\sigma_{j}^{A},\tau_{j}^{A})\right\}  _{j\in
A\setminus\{i\}}}\left[  M_{\tau^{A}}-\int_{0}^{\tau^{A}}c_{i}\left(
\sigma_{i,t}^{A}\right)  dt\right]  .
\]
An \textit{equilibrium} is a profile of Markov strategies satisfying the
assumptions above and constituting best responses for all agents.





\section{Equilibrium Team Search\label{Equilibrium}}

In this section, we characterize the outcomes of team search. We describe the equilibrium search scopes and stopping
boundaries. We also identify the sequencing of agents' search termination, and the patterns of equilibrium exit waves. 

\subsection{Equilibrium Characterization}

Given our restriction on agents' strategies, it follows that any alliance $A$ gets
smaller at the minimal stopping time of its members. That is, the time
$\tau^{A}$ at which the first members of $A$ stop search 
is given by $\tau^{A}=\min_{i\in A}\tau_{i}^{A}$. Equivalently,
\[
\tau^{A}=\inf\{t\geq0:X_{t}=\max_{i\in A}g_{i}^{A}(M_{t})\}\text{.}
\]
Since agents use continuous stopping boundaries, we can write
\[
\tau^{A}=\inf\{t\geq0:X_{t}=g^{A}(M_{t})\},
\]
where $g^{A}(M_t)\equiv\max_{i\in A}g_{i}^{A}(M_{t})$ is continuous.

We start by identifying equilibrium search scopes.
Individual search scopes depend only on the active alliance and are constant
as long as no member departs.

\begin{proposition}[Team Search Scope]\label{prop:Team Search Scope}
For any agent
$i$ in an active alliance $A$, equilibrium search scopes are
constant, $\sigma_{i}^{A}(M_{t},X_{t})=\sigma_{i}^{A}$. Whenever interior, search scopes satisfy the system:
\[
\frac{2c_{i}(\sigma_{i}^{A})}{c_{i}^{\prime}(\sigma_{i}^{A})}=\sigma^{A}=\sum_{i\in A}\sigma_{i}^{A}\text{
\ \ \ \ }\forall i\in A\text{.}
\]

\end{proposition}



Why are search scopes constant as long as a certain alliance of agents is active?
The rough intuition is the following. Consider an agent $i$ in an active
alliance $A$. Suppose $i$ believes that all other agents $j$ in the alliance
search with scope $\sigma_{j}^{A}$. When away from agent $i$'s stopping
boundary, agent $i$ can contemplate a small interval of time in which she is
unlikely to hit her stopping boundary. For that small interval, agent $i$ considers the
induced speed of the process: $\left(  \sum_{k\in A}\sigma_{k}^{A}\right)
^{2}$ and the cost she incurs, $c_i(\sigma_{i}^{A})$. Ultimately, the agent aims
at minimizing the cost per speed, or the overall cost to traverse any distance
on the path,
$\frac{c_{i}(\sigma_{i}^{A})}{\left(  \sum_{k\in A}\sigma_{k}^{A}\right)  ^{2}
}=\frac{c_{i}(\sigma_{i}^{A})}{(\sigma^{A})^{2}}\text{.}$
The identity in the proposition reflects the corresponding first-order
condition.

In general, there might be multiple solutions to the system in Proposition \ref{prop:Team Search Scope}, some possibly corresponding to less efficient equilibria. Nonetheless, \cite{fleming2012deterministic} (Theorem 6.4) guarantees that any equilibrium features continuous search scopes within any alliance. Hence, within an active alliance, agents can utilize only one of the solutions.\footnote{The conditions of Theorem 6.4 in \cite{fleming2012deterministic} follow from our assumption that the cost function's second derivative is bounded above zero. For an alliance composed of one individual, there is a unique optimal solution due to concavity of the objective function.} Importantly, it is the \textit{ratio} of costs to marginal costs that govern equilibrium search scopes. In particular, in our setting, teaming up with agents who have both higher costs and marginal costs can be beneficial in terms of externalities.

Whenever interior solutions to the system in Proposition \ref{prop:Team Search Scope} are unique, comparisons of search scopes within various alliances are well defined. Uniqueness of interior solutions is guaranteed when, e.g., all scope costs are log-convex. A direct corollary of Proposition \ref{prop:Team Search Scope} is then the following. 

\begin{corollary}[Search Scope and Alliance Size]\label{corr:searchscopeandassembly}
Suppose costs are log-convex and an interior solution exists for the systems specified in Proposition \ref{prop:Team Search Scope}. As an alliance
shrinks, individual members' search scopes increase, while total search scope
decreases. That is, for any $i,j\in A$, we have $\sigma
_{i}^{A\backslash\{j\}}\geq\sigma_{i}^{A}$ while $\sigma^{A}%
>\sigma^{A\backslash\{j\}}$.
\end{corollary}


The corollary highlights a form of free-riding. Search scope is
substitutable across individuals. The more agents searching, the less each one
searches. Since individual search scopes decrease within an alliance, the total search
scope in any active alliance is smaller than that which would be generated by
the alliance's members searching independently.\footnote{
Any agent receives a higher payoff within an alliance than she would on her own. Indeed, any agent can
emulate her solo-search policy in an alliance and guarantee at least as high a payoff.} 

The corollary indicates that agents departing would never benefit from continuing search on their own, nor 
from switching to search in a smaller alliance than the one they have left. In particular, our assumption that agents who cease search in an alliance reap the benefits from past discoveries rather than pursue further discoveries with other newly-departed agents is without loss of generality.

We now turn to the characterization of equilibrium stopping boundaries. Agents cease their search whenever search results fall by more than a set amount relative to the observed maximum. 
Consequently, the order in which agents terminate their search is fixed and
does not depend on the realized path of search values.

\begin{proposition}[Alliance Stopping Boundary]\label{prop:assemblystoppingboundary}
There exists an
equilibrium such that, for any agent $i$ in any active alliance $A
$,
\[
g_{i}^{A}(M)=M-\frac{(\sigma^{A})^{2}}{2c_{i}(\sigma_{i}^{A})}.
\]
In particular, agent $i\in\arg\min_{j}\frac{(\sigma^{A})^{2}}
{2c_{j}(\sigma_{j}^{A})}$ is the first to stop in any alliance
$A$. Furthermore, given equilibrium search scopes, there is a unique equilibrium in which stopping boundaries are weakly undominated.
\end{proposition}

Stopping boundaries of the form $g(M)=M-d$ are often termed \textit{drawdown stopping boundaries} with \textit{drawdown size} of $d$. In equilibrium, agents stop whenever the gap between the observed maximum and the current observation exceeds their drawdown size, as identified in the proposition.

To glean some intuition for the structure of the equilibrium stopping
boundary, consider some alliance $A$ and suppose all agents believe that
other members of the alliance will continue searching indefinitely with
search scopes given by Proposition \ref{prop:Team Search Scope}. Each individual agent $i$'s optimization
problem then boils down to a solo searcher's optimization, with others' search simply affecting the experienced search costs. Namely, the induced cost of implementing search scope $\sigma$ is $c_{i}(\sigma-\sum_{j\in A,j\neq i}\sigma_{j}^{A})$. Since agent $i$'s optimization problem is identical when observing $X$ and $M$, or $X+k$ and $M+k$ for any arbitrary constant $k$, her stopping boundary must coincide as well and hence takes the form of a drawdown stopping boundary, see \cite{urgunyariv2020} for further details.
Denote the corresponding drawdown size by $d_{i}^{A}$. Suppose $d_{i}^{A}=\min_{j\in A}d_{j}^{A}$. Consider
then another iteration of best responses, where all agents use the drawdown
stopping boundary calculated as above. Agent $i$ would still be best
responding since, from her perspective, others in the alliance would continue
searching for as long as she does. Furthermore, while other agents may want to
alter their stopping boundary, intuitively, none would want to cease search
before agent $i$ since that would contradict their desire to continue
searching for at least as long as agent $i$ in the first place.

This line of argument suggests that, given equilibrium search scopes, the stopping boundary of the first agent
$i$ to terminate search in any alliance $A$ is determined uniquely when focusing on equilibria in which stopping boundaries are weakly undominated.\footnote{The focus on weakly undominated stopping boundaries---given the equilibrium search scopes---allows us to rule out inefficient equilibria that are an artifact of coordination failures, with multiple agents stopping at an earlier time than desired since other alliance members do so.} Multiplicity of equilibria arises from the stopping boundaries of
other agents $j\in A$. Indeed, any agent $j$ who stops strictly after agent
$i$ is indifferent across all stopping boundaries $g_{j}^{A}(\cdot)$ that
satisfy $g_{j}^{A}(M)>g_{i}^{A}(M)$ for all $M$. Naturally, all such choices
of stopping boundaries by agents other than $i$ do not impact when the
alliance first loses some of its members, nor the search scope while it is
fully active. Consequently, equilibrium outcomes are unique.\footnote{Our analysis indicates a link to other cooperative solution concepts in the spirit of the core. At any point in time, were active agents free to form any coalition to pursue search, or cease search, the externalities present in our environment would imply a unique outcome corresponding to the equilibrium outcome we identify.}

\subsection{Equilibrium Exit Waves}

When all agents have the same costs and solutions are interior, equilibrium
takes a simple form. Team members choose identical search scopes, as
determined by Proposition \ref{prop:Team Search Scope}. They also leave in unison---there is only one
exit wave. Proposition \ref{prop:assemblystoppingboundary} suggests that joint departures may occur even when
individual costs differ. 

To see how those happen, consider any active alliance $A$.
Suppose agent $i$ is first to exit: $d_{i}^{A}=\min_{j\in A}d_{j}^{A}$. Let $Z^{1}=\{i\}$.
Now consider the alliance $A \smallsetminus Z^{1}$ resulting from $i
$'s departure. For all remaining agents, there is then a new drawdown that
governs their decision to stop search. These new
drawdowns are $\{d_{j}^{A\smallsetminus Z^{1}}\}_{j\in A\smallsetminus Z^{1}}$. The discrete drop in
overall search scope induced by $i$'s departure may imply that $d_{j}
^{A\smallsetminus Z^{1}}\leq d_{i}^{A}$ for some $j\in A\smallsetminus Z^{1}$. Let $Z^{2}$ correspond to all these agents together with agent $i$. It follows that, as soon as agent $i$ terminates her search, so will all other agents in $Z^{2}$. We can continue this process recursively to identify the clustered exits that occur in equilibrium. Their characterization depends only on the magnitudes of the drawdown sizes identified in Proposition \ref{prop:assemblystoppingboundary}. In particular, they are identified deterministically. Thus,

\begin{corollary}[Equilibrium Exit Waves]\label{corr:exitwaves}
The order of exists is deterministic, while exit times are stochastic.
\end{corollary}

Our description above suggests that one agent leaving
may trigger the departure of multiple agents---a form of snow-balling effect. This implies that targeted
interventions, subsidizing the search of only particular agents, may impact
the entire path of exit waves.

\subsection{Well-ordered Costs\label{WellOrderedCosts}}

We now consider a particular setting, where the identification of exit waves and their comparative statics is particularly simple.

Suppose agents' cost functions are proportional to one another: 
$c=c_{1}\beta_{1}=c_{2}\beta_{2}\dots=c_{N}\beta_{N}$, %
where $\beta_{1}=1<\beta_{2}<...<\beta_{N}$. That is, agent $1$ has the
highest search costs, while agent $N$ has the lowest search costs.

Proposition \ref{prop:Team Search Scope} implies that all agents in an active alliance choose the same
search scope, assuming an interior solution exists. Suppose $\sigma$ denotes
the search scope all agents utilize in the full alliance. It follows that
$\frac{2c(\sigma)}{c^{\prime}(\sigma)}=N\sigma\text{.}$

As $N$ increases, individual search scopes decrease. Let $\hat{N}$ be the
maximal integer such that
$\frac{2c(\underline{\sigma})}{c^{\prime}(\underline{\sigma})}<\hat
{N}\underline{\sigma}\text{.}$
For any $N>\hat{N}$, there is no interior equilibrium. Furthermore, when there
are $\hat{N}$ agents in the team, individual search scopes are initially
roughly at their minimum $\underline{\sigma}$, while overall search scope is
$\hat{N}\underline{\sigma}$.

Agents' search scope changes only when their alliance shrinks. In this special
case, we can pin down the weak order by which agents stop their search without calculating their corresponding drawdown sizes, which greatly simplifies the analysis. Specifically, Proposition \ref{prop:assemblystoppingboundary} implies that agent $N$ exits no sooner than agent $N-1,$ who exits no sooner than agent $N-2$, and so on. In equilibrium, agents with higher costs terminate search earlier. Can non-trivial exit waves occur when agents' costs are strictly ordered?




Consider any active alliance $\{j,\ldots,N\}$. If
\[
d_{j}^{\{j,...,N\}}\geq d_{j+1}^{\{j+1,\ldots, N\}},d_{j+2}^{\{j+2,\ldots,
N\}},d_{j+k}^{\{j+k,\ldots, N\}},
\]
then agents $j,j+1,j+2,...,j+k$ will all terminate their search at the same
time. Figure 1 depicts an example for $N=10$ individuals. In the figure, once
agent $1$ leaves, agents $2$ and $3$ leave as well. Similarly, once agent $4$
leaves, agent $5$ leaves. And so on. Ultimately, the drawdowns that
govern agents' departures correspond to the \textquotedblleft upper
envelope\textquotedblright\ of the graph depicting $d_{j}^{\{j,\ldots, N\}}$
as a function of $j$.%

\begin{figure}
[ptb]
\begin{center}
\includegraphics[width=0.7\columnwidth]
{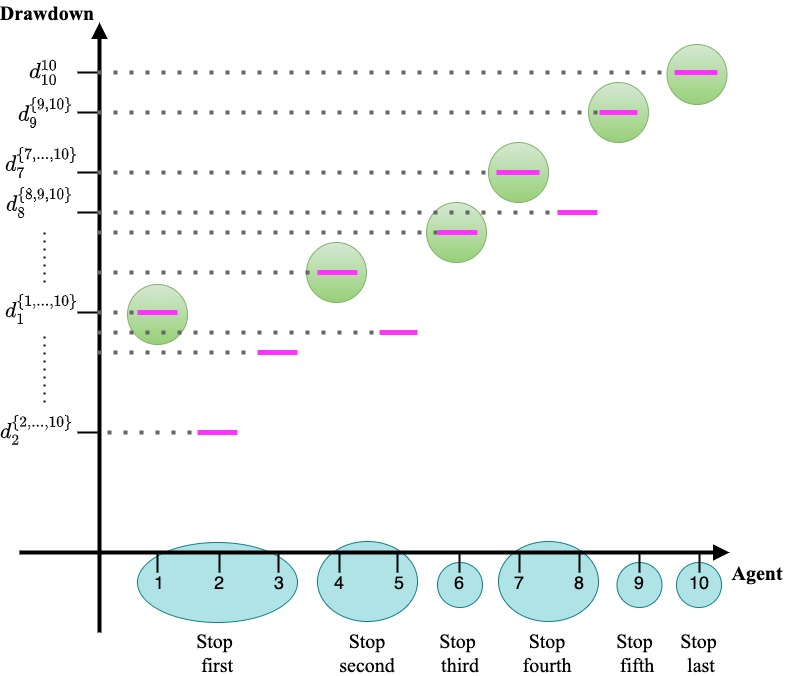}
\caption{Equilibrium exit waves with well-ordered costs}
\end{center}
\end{figure}

Despite agents' costs being strictly ordered, clustered exits are possible. In
fact, when costs are close to one another, all agents might exit at once.
Indeed, from Proposition \ref{prop:assemblystoppingboundary}, $d_{j}^{\{j\ldots N\}}=\frac
{(\sigma^{\{j,\ldots,N\}})^{2}\beta_{j}}{2c(\sigma_{j}^{\{j,\ldots, N\}})}$.
From Corollary \ref{corr:searchscopeandassembly}, $\sigma^{\{j,\ldots, N\}}$ decreases in $j$, while
$\sigma_{j}^{\{j,\ldots,N\}}$ increases in $j$. Therefore, for $\{\beta_{j}\}$
sufficiently close to one another, $d_{1}^{\{1,...,N\}}>d_{2}^{\{2,...,N\}}
>...>d_{N}^{N}$ and all agents exit at once. Naturally, when costs are sufficiently far from one another, agents exit at different points.

A decrease in $\beta_{1}$, keeping $c_{1}\beta_{1}$ and all other
parameters fixed, increases the agent $1$'s search costs and leads to her earlier search termination, potentially too soon for other agents to exit.
Consequently, the number of exit waves weakly increases. In contrast, a decrease in $\beta_{N}$, keeping $c_{N}\beta_{N}$ and all other
parameters fixed, increase agent $N$'s search costs, making her more
inclined to exit when agent $N-1$ does. Consequently, the number of exit waves
weakly decreases.


\section{The Social Planner's Problem\label{Social Planner}}

We now consider a social planner who dictates agents' search scopes and
exit policies to maximize overall utilitarian efficiency of the team. This
analysis highlights the type of inefficiencies that strategic forces in our
joint search process imply. 

\subsection{The Social Objective}

The social planner aims to maximize the agents' expected utilitarian welfare.
The instruments at her disposal are the times at which various agents
exit---the sequence of active alliances---and the search scopes within each
active alliance.

Standard arguments allow us to restrict attention to Markovian policies for
the social planner, see \cite{puterman2014markov}. Formally, we consider a
Markov decision problem in which the state at each date $t$ is
three-dimensional and comprising (i) the set of active agents $A_{t}$, (ii)
the current maximum $M_{t}$, and (iii) the current observed project value
$X_{t}$. The social planner chooses a continuation alliance of agents---a
subset of the current alliance $A_{t}$---and the search scope of each member
in that alliance.

The social planner has two Markovian controls. The first pertains to
the selection of a continuation alliance, and denoted by
$G(M,X,A):\mathbb{R}^{2}\times2^{N}\mapsto2^{A}$. The mapping $G$ determines
the subset of agents continuing the search as a
function of the current state. In particular, if $G(M,X,A)=A$, the current
alliance continues the search. If $\emptyset\neq G(M,X,A)\varsubsetneq A$, the
alliance reduces in size. Whenever $G(M,X,A)=\emptyset$, no agent is left
searching and the search terminates.

The social planner's second control is the profile of search scopes within any
alliance $A$, which can be written as $\sigma_{i}^{A}(M,X):\mathbb{R}^{2}
\mapsto\lbrack\underline{\sigma},\bar{\sigma}]$ for each $i\in A$. 
As before, agents that already exited cannot be induced to choose positive search scope
and do not participate in any future search: exit is irreversible. We
therefore write $\sigma_{i}^A(M,X)=0$ for each $i\notin A$. For any
active alliance $A$, we write:%
\[
\sigma^{A}(M,X)=\sum_{i\in A}\sigma_{i}^{A}(M,X)
\]
and, as a shorthand, we drop the arguments when there is no risk of confusion.

Given these controls, we can now associate a stopping time for each active
alliance $A$. This is the first time at which the alliance shrinks in size.
That is:
\begin{equation}
\tau^{A}=\inf\{t\geq0:G(M_{t},X_{t},A)\neq A\}. \label{zucchini}
\end{equation}
If an alliance $A$ is never reached, we set $\tau^{A}=0$.

Let $\tilde{A}_{t}$ denote the induced process of active alliances. For any
active agent $i$, the time at which her search stops is given by
\[
\tau_{i}=\inf\{t\geq0:i\notin G(M_{t},X_{t},\tilde{A}_{t})\}.
\]
This is the first time at which agent $i$ is not included in an active alliance.

At any time $t$, the welfare of individual $i\in\tilde{A}_{t}$, given the
controls $\{G,\sigma_{i}\}$, is
\[
W_{i}(M_{t},X_{t},\tilde{A}_{t}|\sigma_{i},G)=\mathbb{E}\left[  M_{\tau_{i}%
}-\int_{t}^{\tau_{i}}c_{i}(\sigma_{i,s}^{\tilde{A}_{s}})ds\right]  .
\]
For any $i\notin\tilde{A}_{t}$, we set $W_{i}(M_{t},X_{t},\tilde{A}_{t}%
|\sigma_{i},G)=0$. The social planner's problem is then: 
\begin{align*}
W(M_{t},X_{t},\tilde{A}_{t})  &  =\sup_{\{G,\sigma_{i}\}}\sum_{i}W_{i}
(M_{t},X_{t},\tilde{A}_{t}|\sigma_{i},G)
  =\sup_{\{G,\sigma_{i}\}}\sum_{i}\mathbb{E}\left[  M_{\tau_{i}}-\int
_{t}^{\tau_{i}}c_{i}(\sigma_{i,s}^{\tilde{A}_{s}})ds\right]  .
\end{align*}

We assume that whenever the social planner is indifferent between maintaining
a certain set of agents searching or having them exit, she chooses the latter.

Given a pair of controls ($G,\sigma$), with slight abuse of notation, let $A_{1}=\{1,...,N\}$ denote the first active
alliance, containing all agents.\footnote{We abuse notation by using subscripts to denote the alliance's order in the sequence, rather than time, in order to maintain clarity and simplified notation throughout our analysis.} Using (\ref{zucchini}), let
$A_{2}=\tilde{A}_{\tau^{A_{1}}}$ be the alliance that succeeds the initial alliance, the alliance resulting from the first agents halting their
search. In principle, $A_{2}$ could entail some randomness---depending on the
path observed, different agents may be induced to stop their search. We then
use (\ref{zucchini}) to define $\tau^{A_{2}}$, the (random) time at which the
second set of agents stops search and define $A_{3}=\tilde{A}_{\tau^{A_{2}}}$ as
the (potentially random) resulting alliance. We continue recursively to
establish the (random) time $\tau^{A_{k}}$ at which the $k$'th set of agents
stops search and define $A_{k+1}=\tilde{A}_{\tau^{A_{k}}}$ as the (potentially random)
resulting alliance. Let $K$ denote the (potentially random) number of different active alliances
the social planner utilizes till search terminates for all. 
For any controls $\{G,\sigma_{i}\}$,
we then have a sequence of active alliances $A_{1},A_{2},...,A_{K}$ with
associated stopping times $\tau^{A_{1}},\tau^{A_{2}},...,\tau^{A_{K}}$.

Suppose our team-search problem starts at the state $(M,X,A)$. We set
$\tau^{A_{0}}=0$ and $A_{K+1}=\emptyset$ so that the social planner's problem
can now be written as:
\[
W(M,X,A_{1})=\sup_{\{G,\sigma_{i}\}}\mathbb{E}\left[  \sum_{k=1}^{K}\left(
|A_{k}\setminus A_{k+1}|M_{\tau^{A_{k}}}-\int_{\tau^{A_{k-1}}}^{\tau^{A_{k}}%
}\sum_{i\in A_{k}}c_{i}(\sigma_{i,t}^{A_{k}})dt\right)  \right]  .
\]
Equivalently, we can write the problem recursively starting from any state
$(M,X,A_{k})$:
\[
W(M,X,A_{k})=\sup_{\{G,\sigma_{i}\}}\mathbb{E}\left[  |A_{k}\setminus
A_{k+1}|M_{\tau^{A_{k}}}-\int_{0}^{\tau^{A_{k}}}\sum_{i\in A_{k}}c_{i}%
(\sigma_{i,t}^{A_{k}})dt+W(M_{\tau^{A_{k}}},X_{\tau^{A_{k}}},A_{k+1})\right]
.
\]

Suppose the social planner finds it optimal to halt the search of agent $i$ in
an active alliance $A$ when observing $X$ and $M$. It would then also be
optimal to halt the search of this agent when observing $X^{\prime}$ and $M$
with any $X^{\prime}<X$. Intuitively, the social planner's
solution would be the same were the process shifted by a constant. Therefore,
her choice when observing value $X^{\prime}$ and a maximum $M$ is the same as
when observing $X$ and maximum value $M^{\prime}\equiv M+X-X^{\prime}>M$. As we soon show, search
scopes do not explicitly depend on the achieved maximum. Hence, when
observing $X$ and $M^{\prime}$, were the social planner to continue agent
$i$'s search for a small time interval, the optimal search scopes in the
active alliance would coincide with those she would pick for the same
alliance were search continued when observing $X$ and $M$. 
However, the likelihood of surpassing $M^{\prime}$ at
this small time interval is lower than the likelihood of surpassing $M$.
Furthermore, the social planner could gain $M^{\prime}$ from releasing
agent $i$ with the current observed maximum relative to the lower $M$ she
would get from releasing that agent when observing $X$ and $M$. Thus,
if it is optimal to halt agent $i$'s search when observing $X$ and $M$, it is
also optimal to halt that agent's search when observing $X$ and $M^{\prime}$.

We can therefore write
\[
\tau^{A}=\inf\{t\geq0:X_{t}=g^{A}(M_{t})\},
\]
where $g^{A}(M)=\sup\{X:G(M,X,A)\neq A\}$.\footnote{We implicitly assume,
without loss of generality, that whenever the social planner is indifferent
between halting the search of a subset of agents or continuing their search,
she chooses the former.} This kind of stopping time $\tau^{A}$ is commonly
known as an Az\'{e}ma-Yor stopping time \citep{azema}, with the function
$g^{A}$ defining the corresponding stopping boundary.

For any active alliance $A$, we note that $g^{A}(M)<M$ for all $M$. In other
words, it is never optimal to stop that alliance at any $t$ such that
$M_{t}=X_{t}$. If an alliance searches for a non-trivial amount of time at its
inception, say at time $t_{0}$, it must be that $M_{t_{0}} > X_{t_{0}}$. 
The alliance would then continue searching
jointly even were the planner to observe, at some time $t$, the value $X_{t}$
and recorded maximum of $M_{t}$ with $X_{t}=M_{t}=M_{t_{0}}$. But then the
same should hold when $M_{t}=X_{t}=y,$ with arbitrary $y$; this corresponds to
a shifted problem and does not alter welfare considerations.\footnote{This would not hold were the social planner's objective concave in the maximum observed. 
Concavity introduces new challenges, see
\cite{urgunyariv2020} for a discussion of its impact on single-agent
decisions. Its investigation would be an interesting direction for the
future.}

\subsection{Optimal Team Search}

Our first result illustrates that the social planner chooses constant search
scopes for each active alliance. However, the specification of these
search scopes differs from that dictated by equilibrium.

\begin{proposition}[Optimal Search Scope]\label{prop:welfaresearchisfixed} Search scopes within an alliance are constant and depend only on
the alliance's composition. Furthermore, whenever interior, search scopes satisfy the system:
\[
\frac{2\sum_{i\in A}c_{i}(\sigma_{i}^{A})}{c_{i}^{\prime}(\sigma_{i}^{A})}
=\sum_{i\in A}\sigma_{i}^{A}\;\;\;\forall i\in A\text{.}%
\]

\end{proposition}

The intuition for this result resembles that
provided for equilibrium choices. For any active alliance $A$, the social
planner considers the induced speed of the process, given by $\left(
\sum_{k\in A}\sigma_{k}^{A}\right)  ^{2}$ and the cost she incurs, $\sum_{k\in
A}c_{k}(\sigma_{k}^{A})$. The social planner then aims at minimizing the cost
per speed, or the overall cost to traverse any distance on the path,
\[
\frac{\sum_{k\in A}c_{k}(\sigma_{k}^{A})}{\left(  \sum_{k\in A}\sigma_{k}%
^{A}\right)  ^{2}}=\frac{\sum_{k\in A}c_{k}(\sigma_{k}^{A})}{(\sigma^{A})^{2}%
}\text{.}%
\]
The identity in the proposition reflects the corresponding first-order
condition.

When costs are log-convex, the proposition implies that socially optimal search scopes are higher than
those prescribed in equilibrium. Furthermore, when alliance $A$ is active,
each alliance as a whole searches weakly more under the social planner's
solution. Intuitively, the social planner internalizes the positive
externalities entailed by agents' contributions to the scope of search and
thus specifies greater overall search investments. 
Comparative statics of the socially optimal search scopes resemble those described 
for equilibrium choices in Section \ref{Equilibrium}.

In equilibrium, Corollary \ref{corr:searchscopeandassembly} indicated that, as alliances shrink,
remaining agents increase their search scope. The impacts of agents departing
are quite different in the social planner's solution. As members depart, the
externalities of each remaining agent decline: there are fewer others their
search scope helps. Consequently, the socially optimal search scope of each
individual agent \textit{declines}. That is,

\begin{corollary}[Optimal Scope and Alliance Size]\label{corr:optimalscopeandassembly}
Suppose costs are log-convex and the
equilibrium and social planner's search scopes are interior. Then, in any
alliance, an agent's equilibrium search scope is lower than that agent's
search scope in the social planner's solution. Furthermore, in the social planner's solution, each agent's search scope decreases as her
alliance\ shrinks in size.
\end{corollary}

The sequencing of alliances and their search duration also differ between the
social planner's solution and the corresponding equilibrium: 

\begin{proposition}[Optimal Alliance Sequencing] \label{prop:optimalassemblysequencing}
The socially optimal sequence of alliances is deterministic. For any
deterministic sequence of alliances $A_{1},...,A_{k}$ exerting
optimal search scopes, the socially optimal stopping boundaries are
drawdown stopping boundaries. That is, for each alliance $A_{k}$,
$g^{A_{k}}(M)=M-d_{A_{k}}$with $d_{A_{k}}\in R_{+}$.
Furthermore, the drawdown sizes $\{d_{A_{k}}\}$ exhibit a recursive
structure: for any $k$,%
\[
d_{A_{k}}=\frac{|A_{k}\setminus A_{k+1}|}{2\left(  \frac{\sum_{i\in A_{k}%
}c_{i}(\sigma_{i}^{A_{k}})}{(\sigma^{A_{k}})^{2}}-\frac{\sum_{i\in A_{k+1}%
}c_{i}(\sigma_{i}^{A_{k+1}})}{(\sigma^{A_{k+1}})^{2}}\right)  }.
\]
\end{proposition}

Why does the social planner use drawdown stopping boundaries for various
alliances? Intuitively, for any active alliance $A_{k}$, the social planner
considers the marginal group of agents $A_{k}\setminus A_{k+1}$ whose search
will be terminated next. The relevant marginal added cost per speed for that
group is then%
\[
\frac{\sum_{i\in A_{k}}c_{i}(\sigma_{i}^{A_{k}})}{(\sigma^{A_{k}})^{2}}%
-\frac{\sum_{i\in A_{k+1}}c_{i}(\sigma_{i}^{A_{k+1}})}{(\sigma^{A_{k+1}})^{2}%
}.
\]

Each of these agents would receive the established maximum once they depart,
thereby generating a multiplier of $|A_{k}\setminus A_{k+1}|$ of the maximum
in the social planner's objective. The resulting stopping boundary then
emulates that of a single decision maker, a special case of Proposition \ref{prop:assemblystoppingboundary}, 
with scaled up
returns to each maximum established when the alliance shrinks, and adjusted
costs as above.

To glean some intuition into the deterministic nature of the sequence of
alliances, suppose that the social planner, starting with some active alliance
$A$, proceeds to either alliance $A^{\prime}$ or alliance $A^{\prime\prime}$,
depending on the realized path, with $A^{\prime
},A^{\prime\prime}\subset A$. Following our discussion above, both
transitions---from $A$ to $A^{\prime}$ and from $A$ to $A^{\prime\prime}%
$---are associated with a drawdown stopping boundary, with drawdown sizes of
$d^{\prime}$ and $d^{\prime\prime}$, respectively. If $d^{\prime}%
<d^{\prime\prime}$, starting from alliance $A$, the social planner would
always shrink the alliance to $A^{\prime}$ as the relevant stopping boundary
would always be reached first. Similarly, if $d^{\prime\prime}<d^{\prime}$,
the social planner would always reduce the alliance to $A^{\prime\prime}$. In
other words, different drawdown stopping boundaries never cross one another,
and so the path of alliances is deterministic.

Propositions \ref{prop:welfaresearchisfixed} and \ref{prop:optimalassemblysequencing} suggest that the general structure of efficient search is
similar to that conducted in equilibrium. Agents depart the search process in
a pre-specified order and do so using drawdown stopping boundaries.
Furthermore, within each active alliance, search scopes are constant over
time. Nonetheless, the optimal sequence of active alliances, their
corresponding drawdown sizes, and the search scopes do not generally
coincide with those prescribed by equilibrium.

Certainly, agents who search exert positive externalities on others searching.
Naturally, then, the social planner exploits these externalities by extending
the time individuals spend searching. In fact, the expressions derived for the
optimal and equilibrium alliance drawdown sizes imply directly the following.


\begin{corollary}[Longer Optimal Search]\label{corr:longeroptimalsearch}
Suppose costs are log-convex and the
equilibrium and social planner's search scopes are interior. Consider any alliance that is active on path in both the social planner's solution and in equilibrium. Then, the drawdown chosen by the social planner for that alliance is
weakly larger than the equilibrium drawdown of the same alliance.
\end{corollary}



The results of this section provide some features of the optimal solution.\footnote{In addition, in the Online Appendix, we show a recursive formulation of the social planner's objective---the resulting welfare---in terms of the optimal drawdowns and search scopes.}
However, they do not offer a general characterization of the optimal sequence
of alliances, which is the result of a challenging combinatorial optimization problem---in principle, the planner needs to consider all possible exit patterns, corresponding to ordered partitions of the team.
A sharper characterization requires more structure on the environment's fundamentals. In
the next subsection, we impose such a structure and solve the social planner's
problem completely, illustrating the optimal sequence of alliances and
contrasting it with that emerging in equilibrium.

\subsection{Optimal Team Search with Well-ordered Costs}

Suppose, as in Section \ref{WellOrderedCosts}, that agents' cost functions are
proportional to one another and point-wise ordered:
$c_{1}\beta_{1}=c_{2}\beta_{2}=\dots=c_{N}\beta_{N}$,
where $\beta_{1}=1<\beta_{2}<...<\beta_{N}$.

We start by showing that the social planner uses a similar sequencing of
active alliances to that used in equilibrium.

\begin{lemma}[Optimal and Equilibrium Alliance Sequence]\label{lemma:optimalandequilibriumassemblysequence}
In the
social planner's solution, agent $i$ never terminates search before
agent $j$ if $i>j$. In particular, whenever agent %
$i$ terminates search before agent $j$ in equilibrium, the
social planner terminates agent $i$'s search either with, or before,
agent $j$'s.
\end{lemma}

Intuitively, the social planner optimally terminates the search of agents with
the highest search costs first, so agent $1$'s search is terminated no later
than agent $2$'s search, which is terminated no later than agent $3$'s, etc.
This mimics, \textquotedblleft weakly,\textquotedblright\ the order governed
by equilibrium. Nonetheless, the social planner's sequencing need not echo
that prescribed by equilibrium since clustered exits can differ
dramatically, as we soon show.

It will be useful to introduce the following notation for our characterization
of the socially optimal sequence of alliances. Let
$B_{k}=\{k,k+1,...,N\}\text{ \ \ for all }k=1,...,N$.
Lemma \ref{lemma:optimalandequilibriumassemblysequence} and our equilibrium characterization imply that the optimal sequence
of active alliances has to correspond to a subset of $\{B_{k}\}_{k=1}^{N}$.
This already suggests the computational simplicity well-ordered costs allow.
For instance, instead of considering $2^{N}-1$ alliances that could
conceivably be the last ones active, we need to consider only $N$.

For $B^{\prime}\varsubsetneq B$, we denote by $d_{B\rightarrow B^{\prime}}$
the socially optimal drawdown size associated with alliance $B$, when it is
followed by alliance $B^{\prime}$, as described in Proposition \ref{prop:optimalassemblysequencing}. In
particular, $d_{B\rightarrow\emptyset}$ denotes the optimal drawdown of an
alliance $B$ when it is the last active alliance. We now characterize the optimal sequence of alliances.

\begin{proposition}[Optimal Alliance Sequence with Well-Ordered Costs] \label{prop:optimalassemblysequencewithwellorderedcosts}
The optimal sequence of alliances is identified as follows:

\begin{itemize}
\item There is a unique maximizer of $\{d_{B_{k}\rightarrow\emptyset
}\}_{k=1}^{N}$. Let $L_{1}=\underset{k=1,...,N}{\arg\max}%
d_{B_{k}\rightarrow\emptyset}$. The last active alliance is %
$B_{L_{1}}$, with $L_{1}\leq N$. If $L_{1}=1$, all
agents optimally terminate their search at the same time. Otherwise,

\item There is a unique maximizer of $\{d_{B_{k}\rightarrow B_{L_{1}%
}}\}_{k=1}^{L_{1}-1}$. Let $L_{2}=\underset
{k=1,...,L_{1}-1}{\arg\max}d_{B_{k}\rightarrow B_{L_{1}}}$. The
penultimate active alliance is $B_{L_{2}}$, with $L_{2}<L_{1}%
$. If $L_{2}=1$, there are optimally only two active
alliances: $B_{1}$\textit{\ followed by }$B_{L_{1}}$. Otherwise,

\item Proceed iteratively until reach $L_{n}$, where %
$L_{n}=1$. The socially optimal order of alliances is given by %
$B_{1},B_{L_{1}},\ldots, B_{L_{n-1}}$.
\end{itemize}
\end{proposition}

\begin{figure}
[h]
\includegraphics[width=1\columnwidth
]%
{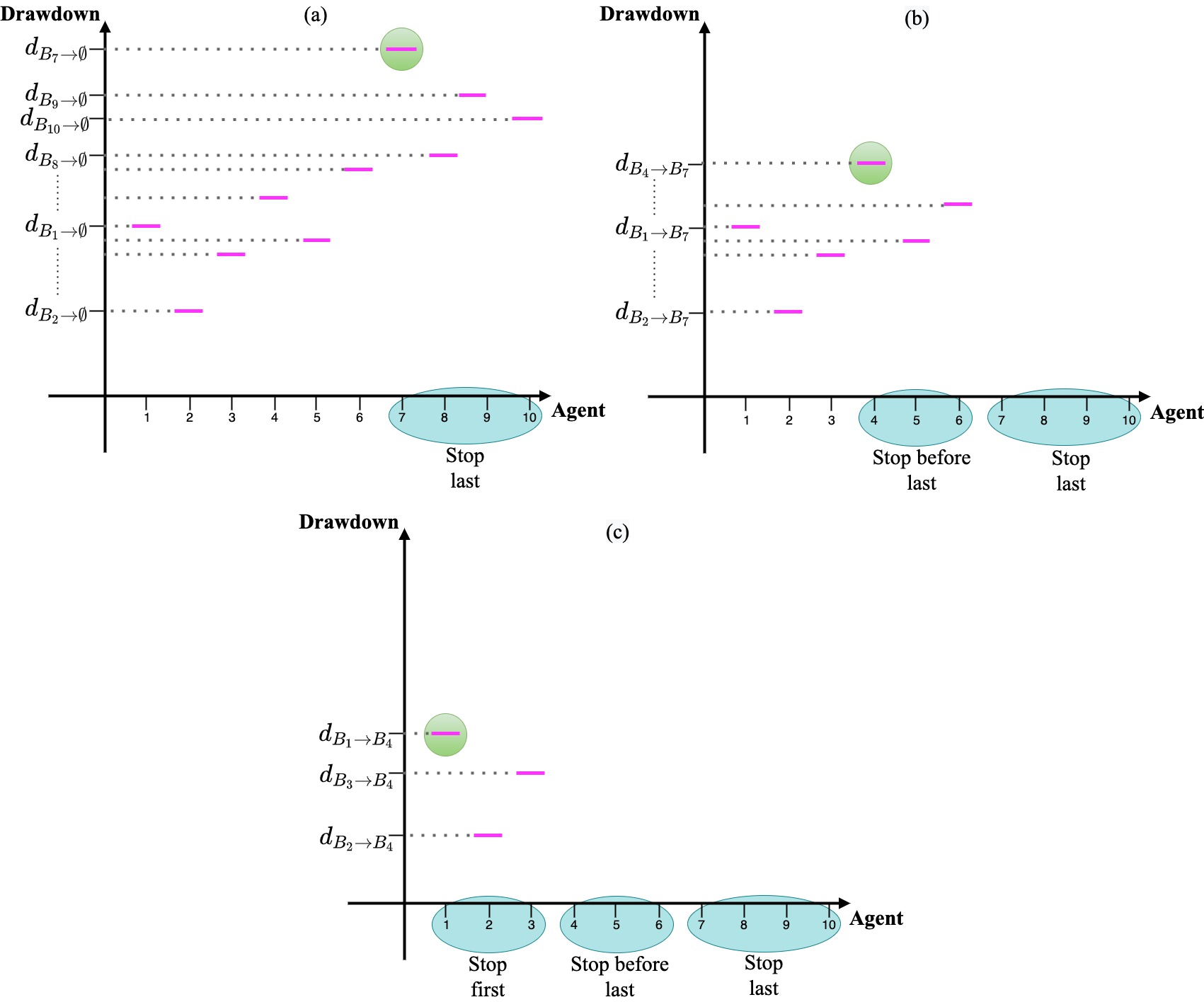}%
\caption{Socially optimal exit waves}%
\end{figure}

The optimal sequence of alliances is constructed recursively. 
Consider first the case in which an alliance's search is terminated
jointly. That is, once search terminates for one of the alliance's members, it
is terminated for all others. Our analysis in the previous section
suggests that, restricted in this way, the social planner would optimally
determine the stopping time using a drawdown stopping boundary. Naturally, any
possible alliance would be associated with a different optimal drawdown size.
Higher drawdown sizes correspond to alliances the planner would prefer to
have searching for longer periods. It is therefore natural to suspect that the
alliance corresponding to the highest such drawdown size is the last active
alliance. Since we already determined that optimal search exits occur in
\textquotedblleft weak\textquotedblright\ order, with agent $i$ never exiting
after agent $i+1$, it suffices to consider drawdown sizes
corresponding to each alliance $B_{k}$.\footnote{As mentioned, this simplifies the computation problem substantially. Instead of considering $2^{N}-1$ alliances, we need to consider only $N$.} 
This allows us to determine the last active
alliance chosen by the social planner, $B_{L_{1}}$, as in panel (a) of Figure 2.

Once $B_{L_{1}}$ is identified, we proceed 
to the penultimate active alliance. Namely, we consider all plausible super-sets of
$B_{L_{1}}$ and assess 
drawdown sizes when the social planner is
constrained to transition directly to $B_{L_{1}}$, see panel (b) of Figure 2. The alliance generating the
maximal such drawdown size is the one the planner would want to keep searching the longest, 
foreseeing her optimal utilization of the next
alliance $B_{L_{1}}$. That is the penultimate alliance. We continue
recursively until reaching the maximal active alliance $B_{1}$, see panel (c) of Figure 2. %


\subsection{Comparing Exit Waves in an Exponential World}

In order to contrast the structure of equilibrium and
socially optimal exit waves, we now consider a particular example. Suppose the
team comprises three agents, $N=3$, and assume cost functions are
exponential and well ordered: $c(\sigma)=c_{1}(\sigma)=e^{b\sigma}=\beta_{2}c_{2}(\sigma)=\beta_{3}c_{3}(\sigma)$,
where $1<\beta_{2}<\beta_{3}$. 
There are four possible exit wave structures: all agents can leave at once; agent $1$ might leave first, followed by the
clustered exit of the lower-cost agents $2$ and $3$; agents $1$ and $2$ might
leave together, followed by agent $3$; or agents may exit at different points.

Figure 3 focuses on the case in which the social planner would cluster all
agents' exits (each tick on the axes corresponds to one unit of the
corresponding multiplier, so that both $\beta_{2}$ and $\beta_{3}$ range from
$0$ to $24$). The figure depicts the different regions of $\beta_{2}$ and
$\beta_{3}$ combinations that generate the four possible structures of
equilibrium exit waves. 
Since $\beta_{3}>\beta_{2}$, all regions are
above the gray $45$-degree diagonal line. We use $\{1,2,3\}$ to denote one
clustered exit wave including all agents; $\{1,2\},\{3\}$ to denote an exit
wave consisting of agents $1$ and $2$, followed by the exit of agent $3$; and
so on.

\begin{figure}[ptb]
\centering
\includegraphics[width=0.8\linewidth]{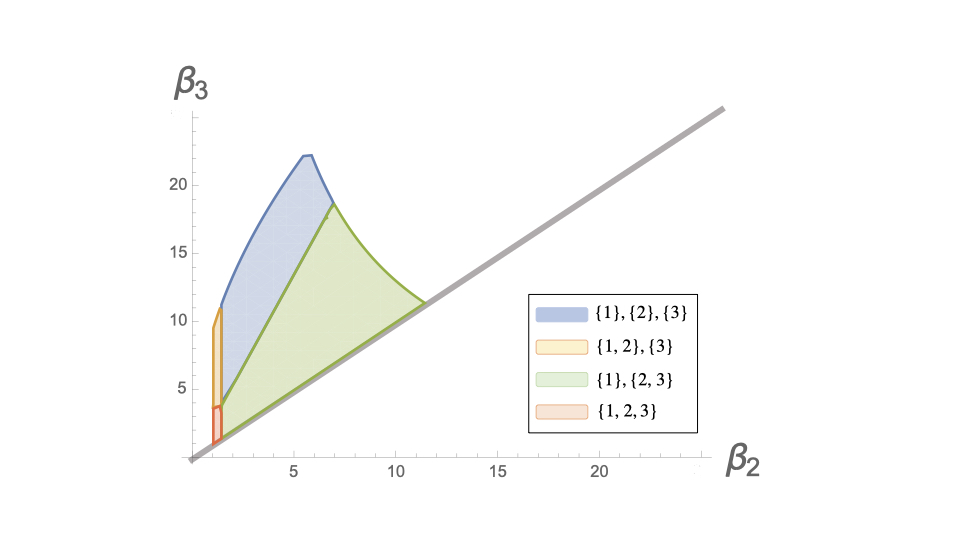}
\caption{Equilibrium exit wave patterns when the optimal policy entails one exit wave including all three agents}
\end{figure}

When the cost multipliers are sufficiently close to one another, agents exit
in unison even in equilibrium. When $\beta_{2}$ is sufficiently close to $1$,
but $\beta_{3}$ is sufficiently higher, agent $3$ has substantially lower
search costs. Since agents $1$ and $2$ do not internalize their externalities
on agent $3$, they prefer to leave early on, generating two exit waves.
Similarly, when $\beta_{2}$ and $\beta_{3}$ are sufficiently high but close to
one another, two exit waves occur in equilibrium. Last, when agents' costs are sufficiently different,
equilibrium dictates agents exiting at different points, resulting in three
exit waves, even when externalities are sufficiently strong so that the social
planner would prefer to have the agents search together till they all exit.
Naturally, for sufficiently high $\beta_{2}$ and $\beta_{3}$, the wedge in
costs is big and even the social planner would prefer to split agents' exits.
The Online Appendix contains detailed characterization of the equilibrium and social planner's solutions and displays similar figures for other exit-wave structures chosen by the social planner.

%

\section{Conclusions and Discussion}

This paper analyzes team search patterns. We show that the equilibrium and
socially optimal search scopes are constant within an alliance. However, as
alliance members depart, individual search scopes increase in equilibrium and
decrease under the optimal policy. We also characterize the deterministic path
of exit waves generated in equilibrium. In particular, even when team members
are fully heterogeneous, clustered exits may occur. The optimal path of exit
waves shares features with the equilibrium path in terms of the
structure of stopping boundaries that govern departures. However, search
externalities naturally prolong optimal search in teams and alter resulting exit waves.

In what follows, we consider two extensions of our model, explicit rewards for innovating early and the utilization of non-Markovian strategies in equilibrium. In the Online Appendix, we also analyze the limitations introduced by a fixed search scope that cannot be altered, and our model's implications for settings with independent search observations.

\subsection{Equilibrium with Penalties for Later Innovations\label{LatePenalties}}

Suppose stopping earlier grants one an advantage. For example, a firm that produces the first product
of its type might capture a market segment that is later more challenging to
capture. Similarly, researchers arguably get additional credit for being the
first to suggest a modeling framework or a measurement technique. 

For simplicity, consider a team of two agents. Assume that the
first agent to stop, say at time $t$, receives $M_{t}$. The second agent to
stop, say at time $s>t$, receives $\alpha M_{s}$, with $\alpha\leq1$. If both
agents stop at the same time $t$, they both receive $M_{t}$.\footnote{The analysis naturally extends to $N$ agents via a decreasing sequence of discounts: $\alpha_{0}=1\geq\alpha_{1}\geq\alpha_{2}\geq...\geq\alpha_{N}$. In addition, one could consider a continuous version of this setup, where the second agent who stops at time $s>t$ receives $M_{t}+\alpha(M_{s}-M_{t})$. That model generates qualitatively similar results, but is more cumbersome to analyze.} As we show in the Online Appendix, the order of exits remains 
deterministic. Furthermore, as long as both agents are searching, the
search scope and the initial stopping boundary are identical to those in our benchmark setting,
where $\alpha=1$. Thus, if there is a unique exit wave when
$\alpha=1$, that is still the case when $\alpha<1$.

Suppose there are two distinct exit waves with $\alpha=1$. Then, there is a
\textit{leader}---the agent who exits early---and a \textit{follower}---the
agent who exits later. The leader's stopping boundary $g_{L}(\cdot)$ remains
her equilibrium stopping boundary regardless of $\alpha$ and is governed by
the drawdown identified in Proposition \ref{prop:assemblystoppingboundary}. 
The follower's stopping boundary, however, may change.

To characterize the follower's stopping boundary, denote the costs of
the leader by $c_{L}(\cdot)$ and those of the follower by $c_{F}(\cdot)$. Let
$\sigma_{L}$ denote the leader's search scope when searching within the full
team, $\sigma_{T}$ denote the total search scope in the
full team, and $\sigma_{F}$ denote the follower's optimal solo search scope.
Similar calculations to those underlying Proposition \ref{prop:assemblystoppingboundary} yield the follower's
stopping boundary $g_{F}(\cdot)$:
\[
g_{F}(M)=%
\begin{cases}
M-\frac{\alpha\sigma_{F}^{2}}{2c_{F}(\sigma_{F})}\text{ if }M<\overline
{M}\text{ and }\frac{\alpha\sigma_{F}^{2}}{2c_{F}(\sigma_{F})}>\frac
{\sigma_{T}^{2}}{2c_{L}(\sigma_{L})},\\
g_{L}(M)\text{ otherwise,}%
\end{cases}
\]
where
\[
\overline{M}=\frac{1}{1-\alpha}\frac{c_{F}(\sigma_{F})}{\sigma_{F}^{2}}\left(
\frac{\alpha\sigma_{F}^{2}}{2c_{F}(\sigma_{F})}-\frac{\sigma_{T}^{2}}%
{2c_{L}(\sigma_{L})}\right)  ^{2}\text{.}%
\]

To glean some intuition, consider the follower's problem after the leader's
departure. The follower faces a similar problem to the individual agent's
problem, with identical search costs and rewards scaled down by $\alpha$. This
case falls within the analysis of \cite{urgunyariv2020}. The
search scope is unaffected by the attenuated rewards, but the drawdown size
is scaled linearly by $\alpha$---as $\alpha$ declines, the rewards from search
become less meaningful, and the follower ceases search more willingly.
Naturally, for sufficiently low $\alpha$, search continuation would not be
worthwhile altogether, regardless of the maximal observation achieved when the
leader exits. That corresponds to the drawdown used by the follower alone,
$\frac{\alpha\sigma_{F}^{2}}{2c_{F}(\sigma_{F})}$, being smaller 
than the full alliance's drawdown, $\frac{\sigma_{T}^{2}}
{2c_{L}(\sigma_{L})}$. In that case, the stopping boundary of the leader
governs the exit of both. In addition, when the maximal observation $M$
achieved when the leader exits is high enough, the loss from leaving at a
later point, $(1-\alpha)M$ is substantial for any $\alpha<1$.\footnote{Specifically, the gain from continuation for the follower is given
by $(1-\alpha)M+(d_{L}-d_{F})^{2}\frac{c_{F}}{\sigma_{F}^{2}}$, where $d_{L}$
and $d_{F}$ are the drawdown sizes for the leader and the follower,
respectively.} For sufficiently high $M$, search continuation would again not
be profitable. 
As $\alpha$ increases, the threshold level
$\overline{M}$ increases. To summarize, for the follower to continue search
after the leader, $\alpha$ needs to be sufficiently high and the current
maximum sufficiently small.

Importantly, when later innovations are penalized, there are no preemption
motives. The main impact is on later innovators, who face weakened incentives
to search. 
Mechanically, larger exit waves occur for a larger set of
parameters. Nonetheless, the main messages of the paper extend directly to
such settings.


\subsection{Non-Markovian Strategies\label{NonMarkovian}}

Our equilibrium analysis restricts attention to Markovian strategies. In our setting, the use of non-Markovian strategies cannot yield the socially optimal solution in general.\footnote{This contrasts insights on collective experimentation, see \cite{horner2020overcoming}.} To see why, consider a team of two agents and suppose the optimal search scope can be implemented in equilibrium---say, when there is only one viable scope, $\underline{\sigma}=\overline{\sigma}$. Our results show that the social planner would like agents to search for a longer time than the (Markovian) equilibrium we identify would prescribe. Suppose agent 1 is the first to exit in such an equilibrium, where stopping strategies are not weakly dominated given the search scopes. As long as agent 2 is searching, agent 1 has a unique best response. 
She would like to use a drawdown size $d_1$, while the social planner would like her to use a drawdown size $d’_1 > d_1$. However, regardless of the space of strategies, there is no way to punish agent $1$ for leaving early, and no way to 
foretell that she will do so. A full analysis of equilibria in non-Markovian strategies is left for the future.

\appendix
\section{Appendix}
  \setcounter{lemma}{0}

Corollary proofs are immediate and, for completeness, available in the Online Appendix. In what follows, we provide proofs for the paper's main results.

\subsection{Proofs for Equilibrium Team Search}


First, we note a useful lemma, commonly known as ``reflection on the diagonal''. This lemma allows us to omit the partial derivatives pertaining to $M$ in the control problem in the various Hamilton-Jacobi-Bellman (HJB) equations that we will derive. Proofs of this result can be found in various sources, including \cite{dubins1994optimal}, \cite{urgunyariv2020} and \cite{peskir1998optimal} among others and hence omitted.
\begin{lemma}\label{lemma:onedimdown}
The infinitesimal generator of the two dimensional process $Z=(M,X)$ satisfies the following:
\begin{enumerate}
\item If $M_{t}>X_{t}$, then $\mathcal{A}_{Z}^{\sigma_{t}}%
=\mathcal{A}_{X}^{\sigma_{t}}=\frac{1}{2}(\sigma_{t})^{2}\frac{\partial^{2}}{\partial
X^{2}}$.
\item If $M_{t}=X_{t}$, then $\frac{\partial V}{\partial
M}=0$.
\end{enumerate}
\end{lemma}

\begin{proof}
[Proof of Proposition \ref{prop:Team Search Scope}]For any agent $i$ in an alliance $A$, the value
function takes the following form
\begin{equation}
V_{i}^{A}(M,X)=\mathbb{E}\left[  \int_{0}^{\tau^{A}}c_{i}(\sigma
_{i,t}|M,X)dt+\mathbb{E}\left[V_{i}^{A}(M_{\tau^{A}},g^{A}(M_{\tau^{A}})\right]\right]  ,
\label{The Claw}
\end{equation}
where
\[
\mathbb{E}\left[  V_{i}^{A}(M_{\tau^{A}},g^{A}(M_{\tau^{A}})\right]  =
\begin{cases}
M_{\tau^{A}}\text{ if }g_{i}^{A}(M_{\tau^{A}})=g^{A}(M_{\tau^{A}})\\
V_{i}^{A\setminus\{j\}}\text{ if }g_{j}^{A}(M_{\tau^{A}})=g^{A}(M_{\tau^{A}
})>g_{i}^{A}(M_{\tau^{A}}),j\in A
\end{cases}
.
\]
In words, with Markov strategies, agent $i$'s expected value is derived from
two components: the cost accrued until her alliance shrinks, and the
continuation value once that happens. If the alliance shrinks with agent $i$'s
departure, her continuation value is simply the maximum value when she exits.

For a given observed maximum $M$, there are two cases to consider for an active agent $i$ in $A$: either her stopping boundary is the highest within the active alliance, or not. We discuss these in sequence.

Suppose first that $g_{i}^{A}(M)=\max_{j\in A}g_{j}^{A}(M)$. Consider
any observed value $X$ such that $g_{i}^{A}(M)\leq X\leq M$. 
The Green function on the interval $[a,b]$ is defined as follows:
\[
\begin{aligned} G_{a, b}(x, y) &=
\left\{\begin{array}{ll} \frac{(b-x)(y-a)}{b-a} & \text { if } a<y<x<b \\ \frac{(b-y)(x-a)}{b-a} & \text { if } a<x<y<b \end{array}\right. \end{aligned}.
\]
Following standard techniques,
we can write the equilibrium value function of agent $i$ in the following recursive fashion:
\begin{align*}
V_{i}^{A}(M,X)&=M\frac{M-X}{M-g_{i}^{A}(M)}+V_{i}^{A}(M,M)\frac{X-g_{i}^{A}%
(M)}{M-g_{i}(M)}\\
&-\int_{g_{i}^{A}(M)}^{M}G_{g_{i}^{A}(M),M}(x,y)c_{i}%
(\sigma_{i}^{A}(M,y))\frac{2}{(\sigma^{A}(M,y))^{2}}dy.
\end{align*}
Rearranging terms, we get:
\begin{align*}
V_{i}^{A}(M,M)-M&=\frac{M-g_{i}^{A}(M)}{X-g_{i}^{A}(M)}\left[  V_{i}%
^{A}(M,X)-M\right.
\\&\left.+\int_{g_{i}^{A}(M)}^{M}G_{g_{i}^{A}(M),M}(x,y)c_{i}(\sigma_{i}%
^{A}(M,y))\frac{2}{(\sigma^{A}(M,y))^{2}}dy\right].
\end{align*}

Since agent $i$ optimally terminates her search at $g_{i}^{A}(M)$, smooth
pasting must hold at $g_{i}^{A}(M)$. The derivative of the continuation value
as $X\rightarrow g_{i}^{A}(M)$ can be written as $\lim_{X\rightarrow g_{i}%
^{A}(M)}\frac{V_{i}^{A}(M,X)-M}{X-g_{i}^{A}(M)}$. By smooth pasting, this must be
equal the derivative of the value from stopping, $\frac{\partial}{\partial
x}M=0$.

Consider then the above equality for $V_{i}^{A}(M,M)$. Taking the limit as
$X\rightarrow g_{i}^{A}(M)$, 
\[
V_{i}^{A}(M,M)=M+\int_{g_{i}(M)}^{M}(M-y)c_{i}(\sigma_{i}^{A}(y))\frac
{2}{(\sigma^{A}(y))^{2}}dy.
\]
This, in turn, implies that
\[
V_{i}^{A}(M,X)=M+\int_{g_{i}^{A}(M)}^{X}(X-y)c_{i}(\sigma_{i}^{A}(y))\frac
{2}{(\sigma^{A}(y))^{2}}dy.
\]
Now, taking the second derivative with respect to $X$ and simplifying yields:
\begin{align*}
\frac{\partial^2 V_i^A(M,X)}{\partial X^2} = \frac{2c_i(\sigma_{i}^A(M,X))}{(\sigma_i^A(M,X))^2}.
\end{align*}
Plugging this back into the HJB for agent $i$ and simplifying further yields:
\begin{align*}
\frac{2 c_i(\sigma_i^A(M,X))}{c'_i(\sigma_i^A(M,X))}=\sigma_i^A(M,X) + \sum_{j \in A \setminus \{i\}}\sigma_j^A(M,X).
\end{align*}

Suppose now that $i$ does not have the highest stopping boundary: $\max_{k\in A}g_{k}^{A}(M) > g_{i}^{A}(M)$. Let $F(M)= \{ j \in N :\max_{k\in A}g_{k}^{A}(M)=g_{j}^{A}(M)\}$. Choose an arbitrary agent $j \in F(M)$. As above, we can write the continuation payoff of $i$ as follows:
\begin{align*}
V_{i}^{A}(M,X)&=V_i^{A }(M,g_i^A(M))  \frac{M-X}{M-g_{j}^{A}(M)}+V_{i}^{A}(M,M)\frac{X-g_{j}^{A}
(M)}{M-g_{j}(M)}\\&-\int_{g_{j}^{A}(M)}^{M}G_{g_{j}^{A}(M),M}(X,y)c_{i}
(\sigma_{i}^{A}(M,y))\frac{2}{(\sigma^{A}(M,y))^{2}}dy.
\end{align*}
Rearranging terms, we get:
\begin{align*}
V_{i}^{A}(M,M)-V_i^{A}(M,g_j^A(M))&=\frac{M-g_{j}^{A}(M)}{X-g_{j}^{A}(M)}\left[  V_{i}
^{A}(M,X)-V_i^{A}(M,g_j^A(M))\right.\\&+\left.\int_{g_{j}^{A}(M)}^{M}G_{g_{j}^{A}(M),M}(X,y)c_{i}(\sigma_{i}
^{A}(M,y))\frac{2}{(\sigma^{A}(M,y))^{2}}dy\right].
\end{align*}
Again, taking the limit as $X \rightarrow g(M)$ from above, and letting $\frac{\partial_D V_{i}^{A}(M,g(M))}{\partial X}$ denote the upper Dini derivative of $V_{i}^{A}(M,g(M))$ at $g(M)$, we have:\footnote{Since $\sigma_i$'s are bounded, $V$ is Lipschitz, hence the Dini derivative is finite.}
\begin{align*}
V_{i}^{A}(M,M)&=\frac{\partial_D V_{i}^{A}(M,g(M))}{\partial X}(M-g(M))+V_i^{A}(M,g_j^A(M))\\&+\int_{g_{j}(M)}^{M}(M-y)c_{i}(\sigma_{i}^{A}(M,y))\frac
{2}{(\sigma^{A}(M,y))^{2}}dy.
\end{align*}
Plugging this identity in $V_i^A(M,X)$'s expression and taking the second derivative:
\begin{align*}
\frac{\partial^2 V_i^A(M,X)}{\partial X^2} = \frac{2c_i(\sigma_{i}^A(M,X))}{(\sigma_i^A(M,X))^2}.
\end{align*}
Plugging this back into the HJB for agent $i$ and simplifying further generates:
\begin{align*}
\frac{2 c_i(\sigma_i^A(M,X))}{c'_i(\sigma_i^A(M,X))}=\sigma_i^A(M,X) + \sum_{k \in A \setminus \{i\}}\sigma_k^A(M,X).
\end{align*}
Our assumption that the cost function's second derivative is bounded above zero implies that the conditions of Theorem 6.4 of \cite{fleming2012deterministic} are satisfied. Hence, any equilibrium features continuous search scopes within any alliance. In particular, within an active alliance, agents can utilize only one of the solutions for the system above.
\end{proof}

\begin{proof}[Proof of Proposition \ref{prop:assemblystoppingboundary}] The statement of Proposition \ref{prop:assemblystoppingboundary} is a
combination of the following claims.

\begin{claim}
For any given alliance $A$ with $i\in A$\textit{,} if $g_{i}^{A}(M^{\ast
})=\max_{j\in A}g_{j}^{A}(M^{\ast})$ for some $M^{\ast}$, then $g_{i}%
^{A}(M)=\max_{j\in A}g_{j}^{A}(M)$ for all $M$.
\end{claim}

\begin{proof}[Proof of Claim]
The proof of the claim relies on the following lemma.

\begin{lemma}
\label{lemma:drawdowninteam} Suppose agent $i\in A$ has the highest stopping
boundary at a given observed $M,X$. Then $g_{i}^{A}(M)$ is a drawdown stopping boundary.
\end{lemma}

\begin{proof}[Proof of Lemma \ref{lemma:drawdowninteam}]
Suppose $\max_{j\in A}g_{j}^{A}(M)=g_{i}^{A}(M)$. As shown in the proof of Proposition \ref{prop:Team Search Scope}, we have
\[
V_{i}^{A}(M,X)=M+\int_{g_{i}^{A}(M)}^{X}(X-y)c_{i}(\sigma_{i}^{A}(M,y))\frac
{2}{(\sigma^{A}(M,y))^{2}}dy.
\]
Furthermore using Proposition \ref{prop:Team Search Scope} we know that $\sigma^A(M,X)=\sigma^A$ for all $M,X$ and $\sigma_i^A(M,X)=\sigma_i^A$ for all $M,X$.

Now, differentiating $V_{i}^{A}(M,X)$ with respect to $M$ and evaluating the
derivative at $X=M$ yields the following ordinary differential equation (ODE)
for $g_{i}^{A}(M)$:
\[
g_{i}^{A}(M)^{\prime}=\frac{(\sigma^{A})^{2}}{2c_{i}(\sigma_{i}^{A})\left(
M-g_{i}^{A}(M)\right)  },
\]
which leads to the following solution:
\[
g_{i}^{A}(M)=M-\frac{(\sigma^{A})^{2}}{2c_{i}(\sigma_{i}^{A})}.
\]
This is a drawdown stopping boundary with drawdown size $d^A_{i}:=\frac
{(\sigma^{A})^{2}}{2c_{i}(\sigma_{i}^{A})}$.
\end{proof}

We can now proceed with the claim's proof. Towards a contradiction, suppose
that $\max_{j\in A}g_{j}^{A}(M)=g_{i}^{A}(M)$ for some $M$. Suppose
$M^{\prime}=\inf_{\hat{M}>M}\{M|i\notin\arg\max_{j\in A}g_{j}^{A}(\hat{M})\}$
and that for some $\varepsilon>0$, for any $\hat{M}\in(M^{\prime
},M^{\prime}+\varepsilon),$ for some $k\neq i$, we have $\max_{j\in A}%
g_{j}^{A}(M^{\prime})=g_{k}^{A}(M^{\prime})>g_{i}^{A}(M^{\prime})$. From
continuity of the stopping boundary and Lemma \ref{lemma:drawdowninteam},
\[
g_{i}^{A}(M)=M-\frac{(\sigma^{A})^{2}}{2c_{i}(\sigma_{i}^{A})}\text{ \ \ \ and
\ \ \ }g_{k}^{A}(M^{\prime})=M^{\prime}-\frac{(\sigma^{A})^{2}}{2c_{k}%
(\sigma_{k}^{A})}.
\]
Our choice of $i$ and $k$ yields $\frac{(\sigma^{A})^{2}%
}{2c_{i}(\sigma_{i}^{A})}\leq\frac{(\sigma^{A})^{2}}{2c_{k}(\sigma_{k}^{A})}$
and $\frac{(\sigma^{A})^{2}}{2c_{i}(\sigma_{i}^{A})}>\frac{(\sigma^{A})^{2}%
}{2c_{k}(\sigma_{k}^{A})}$, in contradiction.
\end{proof}

\begin{claim}
Suppose that for some $i$ in an active alliance of $A$, $\frac{(\sigma
^{A})^{2}}{2c_{i}(\sigma_{i}^{A})}\leq\frac{(\sigma^{A})^{2}}{2c_{j}%
(\sigma_{j}^{A})}$ for all $j\in A$. Then $i$ is the first to exit alliance
$A$.\footnote{If there are multiple agents who satisfy the condition, all
exhibiting the same drawdown size, they all exit jointly, weakly before
others.}
\end{claim}

\begin{proof}[Proof of Claim]
Suppose $\frac{(\sigma^{A})^{2}}{2c_{i}(\sigma_{i}^{A})}\leq\frac{(\sigma
^{A})^{2}}{2c_{j}(\sigma_{j}^{A})}$ for all $j\in A$ but that agent $i$ is not
one of the first agents to exit from alliance $A$ for some path of observed
values. For that path, agent $i$ ceases her search when active at a smaller
alliance $A\setminus K$. Without loss of generality, suppose agent $j $ exits
alliance $A$ first (if there are multiple such agents, pick any) when
observing $M$ and $X$. From Lemma \ref{lemma:drawdowninteam}, agent $j$'s
stopping boundary is characterized by a drawdown. However, from the Claim's
restriction,%
\[
M-\frac{(\sigma^{A})^{2}}{2c_{j}(\sigma_{j}^{A})}\leq M-\frac{(\sigma^{A}%
)^{2}}{2c_{i}(\sigma_{i}^{A})}.
\]


For each $k \in A$, the stopping boundary $g_k^A(M)=M-\frac{(\sigma^{A})^{2}}{2c_{k}(\sigma_{k}^{A})}$ is identified by value matching and smooth pasting. In particular, we have $V_k^A(M,g_k^A(M))=M$. If $g_i^A(M) > g_j^A(M)$, this implies that $V_i^A(M,g_j^A(M)+\epsilon)<M $ for $0<\epsilon< \frac{(\sigma^{A})^{2}}{2c_{j}(\sigma_{j}^{A})}-\frac{(\sigma^{A})^{2}}{2c_{i}(\sigma_{i}^{A})}$. Therefore, agent $i$ would prefer to stop strictly before agent $j$.
\end{proof}

The two claims and Lemma \ref{lemma:drawdowninteam}'s characterization yield the proposition's proof.


\end{proof}

\subsection{Proofs for the Social Planner's Solution}

\begin{proof}[Proof of Proposition \ref{prop:welfaresearchisfixed}] Let $\left\{  \sigma_{i}^{A}(M,X,A)\right\}  $ and $G(M,X,A)$ correspond to a
solution to the social planner's problem. Consider any alliance $A_{k}$ at
some observed values and let $A_{k+1}$ denote the potentially empty random
alliance dictated by this optimal solution. Optimality implies that the
induced search scopes with $A_{k}$ should solve:
\[
\sup_{\{\sigma_{i,t}\}_{i\in A_{k}}}\mathbb{E}\left[  |A_{k}\setminus
A_{k+1}|M_{\tau^{A_{k}}}-\int_{0}^{\tau^{A_{k}}}\sum_{i\in A_{k}}c_{i}%
(\sigma_{i,t}^{A_{k}})dt\right]  .
\]
Following similar steps to the proof of proposition \ref{prop:Team Search Scope} in the equilibrium analysis, the continuation HJB for the social planner can be written as: 
\[
\frac{\partial^{2}W(M,X,A_{k})}{\partial X^{2}}=\frac{\partial^{2} \mathbb{E
}\left[  \int_{0}^{\tau^{A_{k}}}\sum_{i\in A_{k}}c_{i}(\sigma_{i}^{A_{k}%
}(X))dt|M,X\right]  }{\partial X^{2}}.
\]
\bigskip It follows that%
\[
\frac{2\sum_{i\in A_{k}}c_{i}(\sigma_{i}^{A_{k}}(X))}{\sum_{i\in A_{k}}%
\sigma_{i}^{A_{k}}(X)}=c_{j}^{\prime}(\sigma_{i}^{A_{k}}(X))\quad\forall j\in
A_{k}.
\]

Since there is no direct dependence on $X$ on either side, optimal search scopes are independent of
observed values and constant over time for each active alliance.

\end{proof}


\begin{proof}[Proof of Proposition \ref{prop:optimalassemblysequencing}]
The proof follows from two lemmas:

\begin{lemma}
\label{lemma:welfarenonrandomdrawdown} If the set of agents exitting an
alliance is independent of the observed path, each alliance has a
stopping boundary identified by a drawdown size $d_{A_{k}}$.
\end{lemma}

\begin{proof}[Proof of Lemma \ref{lemma:welfarenonrandomdrawdown}]
Let $A_{K}$ be the final alliance in the social planner's problem, with
cardinality $|A_{K}|$. The social planner's problem when left with alliance
$A_{K}$, and observing maximum $M$ and current value $X$, takes the following
form:%
\[
W_{K}(M,X)=\sup_{\tau^{K},\{\sigma_{i}\}_{\imath\in A_{K}}}\mathbb{E}\left[
|A_{K}|M_{\tau^{K}}-\int_{0}^{\tau^{K}}\sum_{i\in A_{K}}c_{i}(\sigma
_{i}^{A_{K}})dt\text{ \ }|\text{ \ }M,X\right]  .
\]
This is tantamount to a single-searcher problem, where search rewards are
scaled by $|A_{K}|$.\ From \cite{urgunyariv2020}, the stopping boundary is
given by:
\[
g^{A_{K}}(M)=M-d_{A_{k}},
\]
where $d_{A_{k}}=\frac{|A_{K}|\sigma^{A_{K}}}{2\sum_{i\in A_{K}}c_{i}%
(\sigma_{i}^{A_{K}})}$.

Consider the social planner's problem when the penultimate alliance $A_{K-1}$
is active and the observed maximum and value are $M$ and $X$, respectively:%
\begin{align*}
W_{K-1}(M,X)  &  =\sup_{\tau^{K-1},\{\sigma_{i}\}_{\imath\in A_{K-1}}%
}\mathbb{E}\left[  |A_{K-1}\setminus A_{K}|M_{\tau^{K-1}}+W_{K}(M_{\tau^{K-1}%
},g^{A_{K-1}}(M_{\tau^{K-1}}))\text{ \ }|\text{ \ }M,X\right] \\
&  -\mathbb{E}\left[  \int_{0}^{\tau^{K-1}}\sum_{i\in A_{K-1}}c_{i}(\sigma
_{i}^{A_{K-1}})dt\text{ \ }|\text{ \ }M,X\right]  .
\end{align*}
By optimality of the stopping time $\tau^{K-1}$, we have smooth pasting of
$W_{K-1}(M,X)$ and $W_{K}(M,X)$. Therefore,
\begin{align*}
W_{K-1}(M,g^{A_{K-1}}(M))  &  =|A_{K-1}\setminus A_{K}|M+W_{K}(M,g^{A_{K-1}%
}(M)),\\
\frac{\partial W_{K-1}(M,g^{A_{K-1}}(M))}{\partial X}{\large |_{X=g^{A_{k-1}}%
}}  &  {\large =\frac{\partial(|A_{K-1}\setminus A_{K}|M+W_{K}(M,g^{A_{K-1}%
}(M)))}{\partial X}|_{X=g^{A_{K-1}}(M)}}.
\end{align*}
Similar to our equilibrium analysis, and using the notation for the Green
function introduced there, we can write the welfare maximization problem as
\begin{align*}
&W_{K-1}(M,X)=   |A_{K-1}\setminus A_{K}|M+W_{K}(M,g^{A_{K-1}}(M))\frac
{M-X}{M-g^{A_{K-1}}(M)}\\
&+W_{K-1}(M,M)\frac{X-g_{A^{K-1}}(M)}{M-g^{A_{K-1}}%
(M)}  -\int_{g^{A_{K-1}}}^{M}G_{g^{A_{K-1}}(M),M}(X,y)\frac{2\sum_{i\in A_{K-1}%
}c_{i}(\sigma_{i}^{A_{K-1}})}{(\sigma^{A_{K-1}})^{2}}dy.
\end{align*}
Letting $X$ approach $g^{A_{K-1}}(M)$, smooth pasting and rearranging
yields:\footnote{The equality follows from
\[
{\large \frac{\partial(|A_{K-1}\setminus A_{K}|M+W_{K}(M,g^{A_{K-1}}%
(M)))}{\partial X}|_{X=g^{A_{K-1}}(M)}=}\left(  X-g^{A_{K-1}}(M)\right)
\int_{g^{A_{K}}(M)}^{g^{A_{K-1}}(M)}\frac{2\sum_{i\in A_{K}}c_{i}(\sigma
_{i}^{A_{K}})}{(\sigma^{A_{K}})^{2}}dx.
\]
}
\\$
W_{K-1}(M,X)=  |A_{K-1}\setminus A_{K}|M+W_{K}(M,g^{A_{K-1}}(M))$ \\
$+\left(
X-g^{A_{K-1}}(M)\right)  \int_{g^{A_{K}}(M)}^{g^{A_{K-1}}(M)}\frac{2\sum_{i\in
A_{K}}c_{i}(\sigma_{i}^{A_{K}})}{(\sigma^{A_{K}})^{2}}dx  +  \int_{g^{A_{K-1}}(M)}^{X}(X-y)\frac{2\sum_{i\in A_{K-1}}c_{i}(\sigma
_{i}^{A_{K-1}})}{(\sigma^{A_{K-1}})^{2}}dy,
$\\

Using the closed-form representation of $W_{K}$ leads to:
\begin{align*}
&W_{K-1}(M,X)=  |A_{K-1}|M+\frac{1}{2}(g^{A_{K-1}}(M)-g^{A_{K}}(M))^{2}%
\frac{2\sum_{i\in A_{K}}c_{i}(\sigma_{i}^{A_{K}})}{(\sigma^{A_{K}})^{2}}\\
 & +\left(  X-g^{A_{K-1}}(M)\right)  (g^{A_{K-1}}(M)-g^{A_{K}}(M))\frac
{2\sum_{i\in A_{K}}c_{i}(\sigma_{i}^{A_{K}})}{(\sigma^{A_{K}})^{2}}\\
&+   \frac{1}{2}(X-g^{A_{K-1}}(M))^{2}\frac{2\sum_{i\in A_{K-1}}c_{i}%
(\sigma_{i}^{A_{K-1}})}{(\sigma^{A_{K-1}})^{2}}.
\end{align*}
To generate an ODE\ that identifies $g^{A_{K-1}}(M)$, we take the derivative
with respect to $M$ that, evaluated at $X=M$, should equal $0$. After some
algebraic manipulations, this ODE\ take the form%
\[
\frac{dg^{A_{K-1}}(M)}{dM}=\frac{|A_{K-1}\setminus A_{K}|}{2\left(
M-g^{A_{K-1}}(M)\right)  \left(  \frac{\sum_{i\in A_{K-1}}c_{i}(\sigma
_{i}^{A_{K-1}})}{(\sigma^{A_{K-1}})^{2}}-\frac{\sum_{i\in A_{K}}c_{i}%
(\sigma_{i}^{A_{K}})}{(\sigma^{A_{K}})^{2}}\right)  }.
\]

It is straightforward to verify that the unique solution for this ODE
satisfying the value-matching condition takes the form $g^{A_{K-1}%
}(M)=M-d_{A_{K-1}}$, where%
\[
d_{A_{K-1}}=\frac{|A_{K-1}\setminus A_{K}|}{2\left(  \frac{\sum_{i\in A_{K-1}%
}c_{i}(\sigma_{i}^{A_{K-1}})}{(\sigma^{A_{K-1}})^{2}}-\frac{\sum_{i\in A_{K}%
}c_{i}(\sigma_{i}^{A_{K}})}{(\sigma^{A_{K}})^{2}}\right)  }.
\]
In particular, the optimal stopping boundary is a drawdown stopping boundary.

Proceeding inductively, for any alliance $m\leq K$, the continuation value
when $M$ and $X$ are observed can be written as:%
\begin{align*}
W_{m}(M,X)=  &  |A_{m}\setminus A_{m+1}|M+W_{m+1}(M,g^{A_{m}}(M))\\
&  +\left(  X-g^{A_{m}}(M)\right)  \sum_{k=m}^{K-1}\left(  \int_{g^{k+1}%
(M)}^{g^{k}(M)}\frac{2\sum_{i\in A_{k+1}}c_{i}(\sigma_{i}^{A_{k+1}})}%
{(\sigma^{A_{k+1}})^{2}}dx\right) \\
-  &  \int_{g^{A_{m}}(M)}^{X}(X-y)\frac{2\sum_{i\in A_{m}}c_{i}(\sigma
_{i}^{A_{m}})}{(\sigma^{A_{m}})^{2}}dy.
\end{align*}
We can then repeat the steps above to generate an analogous ODE for $g^{A_{m}%
}(M)$ and verify that it is uniquely identified as a drawdown stopping
boundary. Namely, $g^{A_{m}}(M)=M-d_{A_{m}}$, where
\[
d_{A_{m}}=\frac{|A_{m}\setminus A_{m+1}|}{2\left(  \frac{\sum_{i\in A_{m}%
}c_{i}(\sigma_{i}^{A_{m}})}{(\sigma^{A_{m}})^{2}}-\frac{\sum_{i\in A_{m+1}%
}c_{i}(\sigma_{i}^{A_{m+1}})}{(\sigma^{A_{m+1}})^{2}}\right)  }.
\]

\end{proof}

\begin{lemma}\label{lemma:welfaredeterministic}
The set of agents dropping from an alliance is deterministic. That is for any
alliance $A$, for all pairs $(M,X),(M^{\prime},X^{\prime})$ such that
$G(M,X,A)\neq A$, we have $G(M,X,A)=G(M^{\prime},X^{\prime},A)$.
\end{lemma}

\begin{proof}[Proof of Lemma \ref{lemma:welfaredeterministic}]
We prove this result by induction on the size of the initial team $N$,
regardless of the starting values of the maximum and the current value. The claim
follows immediately for $N=1$. In that case, the agent uses a drawdown
stopping boundary and the only way for the singleton alliance to change is for
the agent to terminate her search.

For the inductive step, assume that for any initial team of size $N-1$ or less
the optimal alliance sequence is deterministic. By Lemma
\ref{lemma:welfarenonrandomdrawdown}, each of these alliances is associated
with a drawdown stopping boundary. Let $A_{1}$ be an alliance of size $N$. The
continuation value when $M$ and $X$ are observed is:
\[
W_{1}(M,X)=\mathbb{E}\left[  M_{\tau^{1}}+\max_{A_{2}\subsetneq A_{1}}%
\{|A_{1}\setminus A_{2}|M_{\tau^{1}}+W_{A_{2}}(M_{\tau^{1}},g^{A_{1}}%
(M_{\tau^{1}}))\}-\int_{0}^{\tau^{1}}\sum_{i\in A_{k}}c_{i}(\sigma_{i}^{A_{1}%
})dt\right]  .
\]
Suppose that, for some path, the social planner optimally transitions from
alliance $A_{1}$ to a strictly smaller alliance $A_{2}\neq\emptyset$. In
particular, alliance $A_{2}$ contains fewer than $N$ agents. By the inductive
hypothesis, the sequence that ensues is path independent. We can therefore
write the continuation value as:
\begin{align*}
W_{1}(M,X)=  &  |A_{1}\setminus A_{2}|M+W_{2}(M,g^{A_{1}}(M))\\
&  +\left(  X-g^{A_{1}}(M)\right)  \sum_{m=2}^{K-1}\left(  \int_{g^{A_{m+1}%
}(M)}^{g^{A_{m}}(M)}\frac{2\sum_{i\in A_{m+1}}c_{i}(\sigma_{i}^{A_{m+1}}%
)}{(\sigma^{A_{m+1}})^{2}}dx\right) \\
+  &  \int_{g^{A_{1}}(M)}^{X}(X-y)\frac{2\sum_{i\in A_{1}}c_{i}(\sigma
_{i}^{A_{1}})}{(\sigma^{A_{1}})^{2}}dy.
\end{align*}
As before, this yields an ODE characterizing $g^{A_{1}}(M)$ and a unique
solution of the form $g^{A_{1}}(M)=M-d_{A_{1}}$, where
$d_{A_{1}}=\frac{|A_{1}\setminus A_{2}|}{2\left(  \frac{\sum_{i\in A_{1}}%
c_{i}(\sigma_{i}^{A_{1}})}{(\sigma^{A_{1}})^{2}}-\frac{\sum_{i\in A_{2}}%
c_{i}(\sigma_{i}^{A_{2}})}{(\sigma^{A_{2}})^{2}}\right)  }$.
Towards a contradiction, suppose that at some other path a different alliance
is optimally chosen to follow the full alliance $A_{1}$. Call that alliance
$\hat{A}_{2}\neq A_{2}$. Similar argument would then imply that the stopping
boundary for $A_{1}$ is given by $g^{A_{1}}(M)=M-\hat{d}_{A_{1}}$, where
$\hat{d}_{A_{1}}=\frac{|A_{1}\setminus\hat{A}_{2}|}{2\left(  \frac{\sum_{i\in
A_{1}}c_{i}(\sigma_{i}^{A_{1}})}{(\sigma^{A_{1}})^{2}}-\frac{\sum_{i\in\hat
{A}_{2}}c_{i}(\sigma_{i}^{\hat{A}_{2}})}{(\sigma^{\hat{A}_{2}})^{2}}\right)
}$.
Suppose $d_{A_{1}}=d_{\hat{A}_{1}}$. Without loss of generality, assume
$\sigma^{A_{2}}\geq\sigma^{\hat{A}_{2}}$. We cannot have $\hat{A}_{2}\subseteq
A_{2}$. Indeed, if that were the case, the social planner would be indifferent
between keeping the agents in $A_{2}\setminus\hat{A}_{2}$ at points at which
either $\hat{A}_{2}\ $or $A_{2}$ are chosen to continue. However, our
tie-breaking rule implies that such indifferences are broken in favor of
stopping; that is, in such cases, the smaller set $\hat{A}_{2}$ would be
chosen. Therefore, $\hat{A}_{2}\nsubseteqq A_{2}$ and there exists an agent
$i\in\hat{A}_{2}\setminus A_{2}$. Suppose that whenever the social planner
transitions from $A_{1}$ to $A_{2}$, she instead transitions to $A_{2}%
\cup\{i\}$, maintaining the search scopes of members of $A_{2}$ as before and
having agent $i$ search with the lowest scope \underline{$\sigma$} for a
sufficiently small interval of time. In that interval of time, everyone in
$A_{2}$ benefits. When alliance $\hat{A}_{2}$ is picked, agent $i $ uses at
least as high a search scope, while benefitting from lower overall search
scope. In particular, agent $i$ benefits as well from this change.

Suppose $d_{A_{1}}\neq d_{\hat{A}_{1}}$. In this case, the two stopping
boundaries identified above, $M-d_{A_{1}}$ and $M-d_{\hat{A}_{1}}$ never
intersect, in contradiction.
\end{proof}
Combining the two lemmas leads to the conclusion of the proposition.
\end{proof}

\subsection{Proofs for Optimal Sequencing with Well-ordered Costs}

\begin{proof}[Proof of Lemma \ref{lemma:optimalandequilibriumassemblysequence}]
As introduced in the proof of Corollary \ref{corr:optimalscopeandassembly}, we use here the
superscripts $eq$ and $sp$ to denote the equilibrium and social planner's
solution, respectively. When costs are well-ordered, in equilibrium, in any
alliance, all agents utilize the same search scope. In particular, for any
active alliance $A$ and any $i,j\in A$, we have $\sigma_{i}^{A,eq}=\sigma
_{j}^{A,eq}$. This implies that, in equilibrium, each agent $k$ exits no later
than agent $k-1$, for all $k=2,...,N$. Indeed, in any active alliance $A$, the
equilibrium stopping boundary is governed by drawdown size
\[
d_{A}^{eq}=\min_{i\in A}\frac{(\sigma^{A,eq})^{2}}{2c_{i}(\sigma_{i}^{A,eq}%
)}=\max_{i\in A} \frac{(\sigma^{A,eq})^{2}}{2c_{i}(\sigma_{i}^{A,eq})}.
\]
Suppose, towards a contradiction, that there exists a pair $i,j$ such that
$i>j$, so that $\beta_{i}>\beta_{j}$, and the social planner has agent $i$
terminate her search strictly before agent $j$. There are then two distinct
alliances in the social planner's solution, $A_{k}$ and $A_{m},$ with $k<m$,
where $i,j\in A_{k}$ but $i\notin A_{k+1}$ and $j\in A_{m}$ but $j\notin
A_{m+1}$.

As we showed, the social planner's solution associates a drawdown stopping
boundary with each alliance. Denote the corresponding drawdown sizes
$d_{k}^{sp}$ and $d_{m}^{sp}$ for $A_{k}$ and $A_{m}$, respectively. Suppose
that, instead, the social planner swaps the exits of agents $i$ and $j$,
exiting agent $j$ from $A_{k}$ whenever agent $i$ was to cease her search and
exit from $A_{k}$ and exiting agent $i$ from $A_{m}$ whenever agent $j$ was to
cease her search and exit from $A_{m}$. Furthermore, the social planner can
have agent $i$ use the same search scope as agent $j$ had originally in
the alliances that follow $A_{k}$. The overall search scope in any alliance
does not change after this modification. Consequently, expected search
outcomes are unaltered. However, the overall cost decreases weakly in every
alliance and strictly in all alliances $A_{k+1},...,A_{m}$, contradicting
the optimality of the proposed solution.
\end{proof}




\begin{proof}[Proof of Proposition \ref{prop:optimalassemblysequencewithwellorderedcosts}] Recall that our results so far imply that the social planner can restrict
attention to the choice between deterministic alliance sequences. Furthermore,
given a deterministic sequence of alliances, Lemma
\ref{lemma:welfarenonrandomdrawdown} identifies the optimal drawdown stopping
boundaries associated with that alliance sequence. If the chosen sequence is
suboptimal, some of its associated drawdown sizes might be negative or zero,
implying the corresponding alliance is utilized for no length of time. This
observation helps us to identify the optimal sequence. The proof of
Proposition \ref{prop:optimalassemblysequencewithwellorderedcosts} follows from several lemmas. For any alliance $B_{k}$,
regardless of whether it is on the social planner's optimal alliance sequence,
we denote the optimal overall search scope within the alliance by
$\tilde{\sigma}^{k}$ and the consequent overall search cost within that
alliance by $\tilde{c}^{k}$.\bigskip

\begin{lemma}
\label{lemma: largerdroplargerdrawdown} For any $m,j,k$ such that $m<j<k$, if
the welfare-maximizing sequence is such that $B_{k}$ is preceded by $B_{m} $,
then for any sequence where $B_{k}$ is preceded by $B_{j}$, we have
$d_{B_{m}\rightarrow B_{k}}>d_{B_{j}\rightarrow B_{k}}$.
\end{lemma}

\begin{proof}[Proof of Lemma \ref{lemma: largerdroplargerdrawdown}]
from the characterization of drawdowns in the well-ordered settings,
$d_{B_{m}\rightarrow B_{k}}\neq d_{B_{j}\rightarrow B_{k}}$. Suppose that
$d_{B_{m}\rightarrow B_{k}}<d_{B_{j}\rightarrow B_{k}}$. Since $B_{k}$ is
preceded by $B_{m}$ in the optimal sequence, $d_{B_{m}\rightarrow B_{k}}>0$.
It then follows that $d_{B_{j}\rightarrow B_{k}}>0$. This implies that it
would be beneficial for the planner to have alliance $B_{m}$ first transition
to alliance $B_{j}$, and only then transition to alliance $B_{k}$.
\end{proof}

\begin{lemma}
\label{lemma:absolute to margin} If $m<k$, $d_{B_{m}\rightarrow\emptyset
}>d_{B_{k}\rightarrow\emptyset}$ implies $d_{B_{m}\rightarrow B_{k}}%
>d_{B_{k}\rightarrow\emptyset}$.
\end{lemma}

\begin{proof}[Proof of Lemma \ref{lemma:absolute to margin}]
$d_{B_{m}\rightarrow\emptyset}>d_{B_{k}\rightarrow\emptyset}$ implies%
\begin{align*}
\frac{1}{N-m}\frac{\tilde{c}^{m}}{\tilde{\sigma}^{m}}  &  <\frac{1}{N-k}%
\frac{\tilde{c}^{k}}{\tilde{\sigma}^{k}} \Longrightarrow\frac{1}{k-m}(\frac{\tilde{c}^{m}}{\tilde{\sigma}^{m}}%
-\frac{\tilde{c}^{k}}{\tilde{\sigma}^{k}})   <\frac{1}{N-k}\frac{\tilde
{c}^{k}}{\tilde{\sigma}^{k}},
\end{align*}
illustrating the claim.
\end{proof}

\begin{lemma}
\label{lemma: reversalwithfinal} For any $k$ such that $d_{B_{k}%
\rightarrow\emptyset}>d_{B_{N}\rightarrow\emptyset}>d_{B_{k+1}\rightarrow
\emptyset} $, we have $d_{B_{k} \rightarrow B_{k+1}}>d_{B_{N}\rightarrow\emptyset}$.
\end{lemma}

\begin{proof}[Proof of Lemma \ref{lemma: reversalwithfinal}]
Observe that $d_{B_{k}\rightarrow\emptyset}>d_{B_{N}\rightarrow\emptyset
}>d_{B_{k+1}\rightarrow\emptyset}$ implies
\[
(N-k+1)\frac{\tilde{c}^{N}}{\tilde{\sigma}^{N}}>\frac{\tilde{c}^{k+1}}%
{\tilde{\sigma^{k+1}}}\text{ \ \ \ and \ \ \ }\frac{\tilde{c}^{k}}{\tilde{\sigma}^{k}%
}>(N-k)\frac{\tilde{c}^{N}}{\tilde{\sigma}^{N}}.
\]
Simply summing the inequalities and reorganizing yields the implied statement.
\end{proof}

\begin{lemma}
\label{lemma:orderingmarginandabsolute} If $d_{B_{k}\rightarrow\emptyset
}>d_{B_{k-1} \rightarrow B_{k}}$, then $d_{B_{k}\rightarrow\emptyset}>d_{B_{k-1}%
\rightarrow\emptyset}>d_{B_{k-1} \rightarrow B_{k}}.$
\end{lemma}

\begin{proof}[Proof of Lemma \ref{lemma:orderingmarginandabsolute}]
From the first inequality, $d_{B_{k}\rightarrow\emptyset}>d_{B_{k-1} \rightarrow B_{k}} $,
we have,
\begin{align*}
\frac{1}{N-k}\frac{\tilde{c}^{k}}{\tilde{\sigma}^{k}}  &  <\frac{\tilde
{c}^{k-1}}{\tilde{\sigma}^{k-1}}-\frac{\tilde{c}^{k}}{\tilde{\sigma}^{k}
}\Longrightarrow\frac{N-(k-1)}{N-k}\frac{\tilde{c}^{k}}{\tilde{\sigma}^{k}
}<\frac{\tilde{c}^{k-1}}{\tilde{\sigma}^{k-1}}
\Longrightarrow
d_{B_{k}\rightarrow\emptyset}>d_{B_{k-1}\rightarrow\emptyset}.
\end{align*}
But this inequality implies that\\

\noindent$
\frac{1}{N-k}\frac{\tilde{c}^{k}}{\tilde{\sigma}^{k}}    <\frac{1}%
{N-(k-1)}\frac{\tilde{c}^{k-1}}{\tilde{\sigma}^{k-1}}\Longrightarrow \frac{\tilde{c}^{k-1}}{\tilde{\sigma}^{k-1}}-\frac{\tilde{c}^{k}}%
{\tilde{\sigma}^{k}}    >\frac{1}{(N-(k-1))}\frac{\tilde{c}^{k-1}}%
{\tilde{\sigma}^{k-1}}\Longrightarrow d_{B_{k-1}\rightarrow\emptyset
}>d_{B_{k-1} \rightarrow B_{k}}.$

\end{proof}




\begin{lemma}
\label{lemma:welfarelastcantbelarger} If $k$ satisfies $\max_{j}%
d_{B_{j}\rightarrow\emptyset}=d_{B_{k}\rightarrow\emptyset}$, then any
alliance $B_{l}$ with $l<k$ cannot be the welfare maximizing last alliance.
\end{lemma}

\begin{proof}[Proof of Lemma \ref{lemma:welfarelastcantbelarger}]
Suppose not, so that, form some $l<k$, alliance $B_{l}$ is the last. Since
$B_{k}$ is strictly contained in $B_{l}$, from the characterization of
drawdowns in the well-ordered settings, $d_{B_{k}\rightarrow\emptyset}\neq
d_{B_{l}\rightarrow\emptyset}.$ The social planner would, then, benefit from
transitioning from $B_{l}$ to $B_{k}$ instead of exiting all members of
$B_{l}$ since $d_{B_{k}\rightarrow\emptyset}>d_{B_{l}\rightarrow\emptyset}>0$,
in contradiction.
\end{proof}

\begin{lemma}
\label{lemma:welfarelastcantbesmaller} If $k$ satisfies $\max_{j}%
d_{B_{j}\rightarrow\emptyset}=d_{B_{k}\rightarrow\emptyset}$, then any
alliance $B_{l}$ with $l>k$ cannot be the last.
\end{lemma}

\begin{proof}[Proof of Lemma \ref{lemma:welfarelastcantbesmaller}]
We use induction on the cardinality of the set $B_{k}$. The claim certainly
holds when $\left\vert B_{k}\right\vert =1$, so that $B_{k}=B_{N}=\{N\}$.

For the proof, it is useful to notice that our entire analysis does not hinge
on the range of viable search scopes coinciding across agents. In fact, the
analysis would go through in its entirety if each agent $i$ had an individual
range of scope $[\underline{\sigma}_{i},\overline{\sigma}_{i}]$, as long as
all solutions remain interior.

Assume the statement is true for sets up to cardinality $n$. We show the
statement holds for $\left\vert B_{k}\right\vert =n+1$ (so that $k=N-n+1$). By
Lemma \ref{lemma:welfarelastcantbelarger}, the last alliance cannot be $B_{j}$
with $j<k$. Towards a contradiction, suppose that a smaller set $B_{m}$,
with $m>k$, is the last alliance. From the inductive hypothesis, we must have
$d_{B_{m}\rightarrow\emptyset}>d_{B_{l}\rightarrow\emptyset}$ for all $l>m$,
as otherwise the social planner would benefit by inducing $B_{l}$ to continue
search instead of terminating it for all agents in $B_{m}$.\medskip

Suppose that $m<N$. Consider an equivalent problem, where alliance $B_{m}\ $is
replaced with a single individual $M$ that has cost function $\hat{c}(\cdot)$
defined so that $\hat{c}(\sigma)$ is the minimal overall cost in $B_{m}$
required for implementing an overall search scope $\sigma$. That is, if
\[
\sigma_{m}^{B_{m}},...,\sigma_{N}^{B_{m}}\in\arg\min_{\hat{\sigma}_{j}^{B_{m}%
}\in\lbrack\underline{\sigma},\overline{\sigma}]\text{ \ }\forall j,\sum
_{j=m}^{N}\hat{\sigma}_{j}^{B_{m}}=\sigma}\sum_{j=m}^{N}c(\hat{\sigma}%
_{j}^{B_{m}})\text{,}%
\]
then $\hat{c}(\sigma)=\sum_{j=m}^{N}c(\hat{\sigma}_{j}^{B_{m}})$. Under this
definition, $\sigma\in\lbrack(N-m+1)\underline{\sigma},(N-m+1)\overline
{\sigma}]$.

In the equivalent problem, we have $m$ agents $1,2,...,m-1,M$.\footnote{Our
assumption that all alliance optimally have members using an interior search
scope guarantees that this fictitious agent would choose an interior search
scope as well.} From our construction, in the optimal solution, 
for any $j=1,...,m-1$, the corresponding drawdown size
$d_{\{j,...,M\}\rightarrow\emptyset}$ coincides with the optimally-set
drawdown size $d_{B_{j\rightarrow\emptyset}}$ in our original problem.
Furthermore, $d_{\{M\}\rightarrow\emptyset}$ coincides with
$d_{B_{m\rightarrow\emptyset}}$ in our original problem. Therefore, $\max_{j\in\{1,...,m-1,M\}}d_{\{j,...,M\}\rightarrow
\emptyset}=d_{\{k,...,M\}\rightarrow\emptyset}$. By our induction hypothesis,
$\{j,...,M\}$ with $j>k$ cannot optimally be the last alliance, in
contradiction.\medskip

Suppose now that $m=N$ and, towards a contradiction, assume $B_{N}$ is the
welfare maximizing last alliance. Now consider the sequence of welfare
maximizing alliances $B_{p}$ such that $B_{p}\subset B_{k}$. There are three
cases to consider.\medskip

\textbf{Case 1:} For all $p$ $\in\{k,...,N-1\}$, the alliance $B_{p}$ is part
of the welfare-maximizing sequence. That is, agents terminate their search one
by one starting from $B_{k}$ onwards. Since $B_{N}$ is the last alliance, we
must have that
$d_{B_{N}\rightarrow\emptyset}>d_{B_{N-1}\rightarrow B_{N}}>d_{B_{N-2}%
\rightarrow B_{N-1}}>\ldots>d_{B_{k}\rightarrow B_{k+1}}$.
Applying Lemma \ref{lemma:orderingmarginandabsolute} repeatedly implies that
$d_{B_{N}\rightarrow\emptyset}>d_{B_{N-1}\rightarrow\emptyset}>d_{B_{N-2}%
\rightarrow\emptyset}\ldots>d_{B_{k+1}\rightarrow\emptyset}$.
The assumed maximality of $d_{B_{k}\rightarrow\emptyset}$ implies, in
particular, that $d_{B_{k}\rightarrow\emptyset}>d_{B_{N}\rightarrow\emptyset}$
that, combined with the above, yields $d_{B_{k}\rightarrow\emptyset}%
>d_{B_{N}\rightarrow\emptyset}>d_{B_{k+1}\rightarrow\emptyset} $. By Lemma
\ref{lemma: reversalwithfinal}, we then have that $d_{B_{k}\rightarrow
B_{k+1}}>d_{B_{N}\rightarrow\emptyset}$. It follows that whenever agents in
the active alliance $B_{k}$ optimally stop searching, the social planner would
benefit from halting all agents' search instead of proceeding with
$B_{k+1},B_{k+2},...,B_{N}$, in contradiction.\medskip

\textbf{Case 2:} There does not exist any $p\in\{k,...,N-1\}$ such that
$B_{p}$ is part of the optimal sequence. Thus, the penultimate alliance in the
optimal sequence is $B_{l}$ with $l<k$. Maximality of $d_{B_{k}%
\rightarrow\emptyset}$ implies that $d_{B_{k}\rightarrow\emptyset
}>d_{B_{N}\rightarrow\emptyset}$. By Lemma \ref{lemma:absolute to margin},
$d_{B_{k}\rightarrow B_{N}}>d_{B_{N}\rightarrow\emptyset}$ and by Lemma
\ref{lemma: largerdroplargerdrawdown}, $d_{B_{l}\rightarrow B_{N}}%
>d_{B_{k}\rightarrow B_{N}}>d_{B_{N}\rightarrow\emptyset}$. Thus, whenever
agents in active alliance $B_{l}$ optimally stop searching, the social
planner would benefit from halting all agents' search instead of proceeding
with $B_{N}$, in contradiction.\medskip

\textbf{Case 3:} There exist $p,q\in\{k,...,N-1\}$ such that $B_{p}$ is part
of the optimal sequence but $B_{q}$ is not. Here we have two subcases:

\indent\textbf{Subcase 1:} $B_{N-1}$ is the penultimate alliance. We must have
$d_{B_{N-1}\rightarrow\emptyset}<d_{B_{N}\rightarrow\emptyset}$; otherwise, by
Lemma \ref{lemma:absolute to margin}, we would have $d_{B_{N-1} \rightarrow B_{N}%
}>d_{B_{N}\rightarrow\emptyset}$ and it would be suboptimal to utilize
alliance $B_{N}$ as the last alliance. From the maximality of $d_{B_{k}%
\rightarrow\emptyset}$ and Lemma \ref{lemma: largerdroplargerdrawdown}, for
any $l<k$ such that $B_{l}$ precedes $B_{N-1}$ on the optimal path,
$d_{B_{l}\rightarrow B_{N-1}}>d_{B_{k}\rightarrow B_{N-1}}>d_{B_{N-1}%
\rightarrow\emptyset}$. Finally, $d_{B_{N}\rightarrow\emptyset}>d_{B_{N-1}%
\rightarrow\emptyset}$ implies that
\[
\frac{1}{2}\frac{\tilde{c}^{N-1}}{\tilde{\sigma}^{N-1}}>\frac{\tilde{c}^{N}%
}{\tilde{\sigma}^{N}}\Longrightarrow\frac{1}{2}\frac{\tilde{c}^{N-1}}%
{\tilde{\sigma}^{N-1}}<\frac{\tilde{c}^{N-1}}{\tilde{\sigma}^{N-1}}%
-\frac{\tilde{c}^{N}}{\tilde{\sigma}^{N}}\Longrightarrow d_{B_{N-1}%
\rightarrow\emptyset}>d_{B_{N-1}\rightarrow B_{N}}.
\]
Thus,
$d_{B_{l}\rightarrow B_{N-1}}>d_{B_{k}\rightarrow B_{N-1}}>d_{B_{N-1}%
\rightarrow\emptyset}>d_{B_{N-1}\rightarrow B_{N}}$.
Therefore, whenever agents in the active alliance $B_{l}$ optimally stop
searching, the social planner would benefit from transitioning to $B_{N}$
directly, thereby terminating the search of agent $N-1$ as well, instead of
transitioning to $B_{N-1}$ first, in contradiction.

\indent\textbf{Subcase 2}: The penultimate alliance is $B_{p}$ with
$p\in\{k,...,N-2\}$. We can now emulate the argument above pertaining to the
construction of an equivalent problem in which agents $\{p,...,N-1\}$ are
viewed as one agent with appropriately induced search costs. We can then
consider an equivalent problem with fewer agents to achieve a contradiction
through our induction hypothesis.
\end{proof}

It follows that the last alliance is given by $B_{k}$
with $\max_{j}d_{B_{j}\rightarrow\emptyset}=d_{B_{k}\rightarrow\emptyset}$.

The proofs of the following Lemmas are a consequence of identical arguments to
those in of Lemmas \ref{lemma:welfarelastcantbelarger} and
\ref{lemma:welfarelastcantbesmaller} and are therefore ommitted.

\begin{lemma}
Consider $B_{k}$ where $k$ is such that $d_{B_{k}\rightarrow B_{L_{1}}%
}>d_{B_{j}\rightarrow B_{L_{1}}}$ for all $j<L_{1}$ and $B_{L_{1}}$ as the
last alliance as identified above. Then any alliance with $l<k$ cannot be the
welfare maximizing second to last alliance.
\end{lemma}

\begin{lemma}
Consider $B_{k}$ where $k$ is such that $d_{B_{k}\rightarrow B_{L_{1}}%
}>d_{B_{j}\rightarrow B_{L_{1}}}$ for all $j<L_{1}$ and $B_{L_{1}}$ as the
last alliance as identified above. Then any alliance $B_{l}$ with ${L_{1}%
}>l>k$ cannot be the welfare maximizing second to last alliance.
\end{lemma}

The proof of Proposition \ref{prop:optimalassemblysequencewithwellorderedcosts} then follows. Using the proposition's notation,
$B_{L_{1}}$ is the last alliance on the social planner's optimal path.
Similarly, 
the penultimate alliance is given by
$B_{k}$ where $k$ is such that $d_{B_{k}\rightarrow B_{L_{1}}}>d_{B_{j}%
\rightarrow B_{L_{1}}}$ for all $j<L_{1}$. We can continue recursively to
establish the proposition's claim.
\end{proof}

\bibliographystyle{chicago}
\bibliography{TeamSearch}

\end{document}